\documentclass[aps,twocolumn,superscriptaddress,prl]{revtex4-2}

\usepackage{bm,graphicx,textcomp,amssymb,amsmath,dcolumn,color,xcolor}

\usepackage{tabularx}

\usepackage[utf8]{inputenc}
\usepackage[T1]{fontenc}

\usepackage{notes2bib}

\newcommand{\nn}{\nonumber\\}

\newcommand{\f}[1]{\mbox{\boldmath$#1$}}
\newcommand{\fk}[1]{\mbox{\boldmath$\scriptstyle#1$}}

\newcommand{\ord}{\,{\cal O}}

\newcommand{\ii}{{\rm i}}
\newcommand{\abs}[1]{{\left| #1 \right|}}
\newcommand{\expval}[1]{{\langle #1 \rangle}}

\newcommand{\new}[1]{#1} 

\begin{document}

\title{Kibble-Zurek dynamics in the anisotropic Ising model of the Si(001) surface}

\author{G.~Schaller}
\affiliation{Helmholtz-Zentrum Dresden-Rossendorf, 
Bautzner Landstra{\ss}e 400, 01328 Dresden, Germany,}

\author{F.~Queisser} 
\affiliation{Helmholtz-Zentrum Dresden-Rossendorf, 
Bautzner Landstra{\ss}e 400, 01328 Dresden, Germany,} 

\author{S.P.~Katoorani} 
\affiliation{Helmholtz-Zentrum Dresden-Rossendorf, 
Bautzner Landstra{\ss}e 400, 01328 Dresden, Germany,} 

\author{C.~Brand}
\affiliation{Fakult\"at f\"ur Physik, Universit\"at Duisburg-Essen,
  Lotharstra{\ss}e 1, 47057 Duisburg, Germany,}

\author{C.~Kohlf\"urst} 
\affiliation{Helmholtz-Zentrum Dresden-Rossendorf, 
Bautzner Landstra{\ss}e 400, 01328 Dresden, Germany,} 
 
\author{M.R.~Freeman}
\affiliation{Department of Physics, University of Alberta, 
4-181 Centennial Center for Interdisciplinary Science Edmonton, 
Alberta T6G 2E1, Canada,} 

\author{A.~Hucht}
\affiliation{Fakult\"at f\"ur Physik, Universit\"at Duisburg-Essen,
  Lotharstra{\ss}e 1, 47057 Duisburg, Germany,}

\author{P.~Kratzer}
\affiliation{Fakult\"at f\"ur Physik, Universit\"at Duisburg-Essen,
  Lotharstra{\ss}e 1, 47057 Duisburg, Germany,}

\author{B.~Sothmann}
\affiliation{Fakult\"at f\"ur Physik, Universit\"at Duisburg-Essen,
  Lotharstra{\ss}e 1, 47057 Duisburg, Germany,}
  
\author{M.~Horn-von Hoegen}
\affiliation{Fakult\"at f\"ur Physik, Universit\"at Duisburg-Essen,
  Lotharstra{\ss}e 1, 47057 Duisburg, Germany,}

\author{R.~Sch\"utzhold}
\affiliation{Helmholtz-Zentrum Dresden-Rossendorf, 
Bautzner Landstra{\ss}e 400, 01328 Dresden, Germany,}
\affiliation{Institut f\"ur Theoretische Physik, 
Technische Universit\"at Dresden, 01062 Dresden, Germany,}

\date{\today}

\begin{abstract}
As a simplified description of the non-equilibrium dynamics of buckled
dimers on the Si(001) surface, we consider the anisotropic 2D 
Ising model and study the freezing of spatial correlations during a 
cooling quench across the critical point. 
\new{Depending on the cooling rate,}
we observe a crossover from 1D to 2D behavior.
For rapid cooling, we find effectively 1D behavior in the strongly coupled 
direction, for which we provide an exact analytic solution of the
non-equilibrium dynamics.
For slower cooling rates, we start to see 2D behavior where our numerical 
simulations show an approach to the usual Kibble-Zurek scaling in 2D. 
\end{abstract}

\maketitle

\paragraph{Introduction} 

Von Neumann once~\cite{bak1995a} compared non-equilibrium theory to a theory 
of non-elephants -- indicating the richness and complexity of this field, 
which we are just beginning to understand. 
In view of the diverging response time near the critical point, 
continuous phase transitions are prototypical candidates for observing 
non-equilibrium behavior~\cite{sondhi1997a,sachdev2011}.
A prominent example is the Kibble mechanism describing the formation of 
topological defects during symmetry-breaking phase transitions in the 
early universe~\cite{kibble1976a}.
Later Zurek realized that quite analogous effects should also occur 
in condensed matter such as superfluid helium~\cite{zurek1985a}.
The Kibble-Zurek mechanism has been studied in numerous theoretical
(e.g.,~\cite{kibble1997a,zurek2005a,damski2005a,cincio2007a,dutta2010a,delcampo2010a,
liu2014a,chesler2015a,sonner2015a,silvi2016a,jaschke2017a,dora2019a,
puebla2019a,ulcakar2020a,hodsagi2020a,oshiyama2020a,reichhardt2022a,bacsi2023a,weitzel2024a})
and experimental investigations (e.g.~\cite{ruutu1996a,eltsov2000a,ulm2013a,lamporesi2013a,
deutschlaender2015a,gong2016a,beugnon2017a,qiu2020a,rysti2021a,du2023a,li2023a}).
An important point is the transition from adiabatic evolution to 
non-equilibrium behavior (such as freezing) when approaching or 
traversing the critical point. 
Apart from the original idea of creating topological defects, 
the general mechanism can also be applied to
the frozen domain 
structure in symmetry-breaking phase transitions induced by the  
critical slowing down.

In this Letter, we
study the anisotropic Ising model in two 
spatial dimensions~\cite{schultz1964a,mccoy1973,baxter1989,neto2006a,
HobrechtHucht:SciPostPhys7,HobrechtHucht:SciPostPhys8,hucht2021a,zandvliet2023a}
with special emphasis on possible differences 
between the two directions. 
Apart from advancing our fundamental understanding, these investigations
are also motivated by the fact that the buckling dynamics of dimers on 
the Si(001) surface can be described by the anisotropic 2D Ising 
model~\cite{Saxena:SurfSci160.618,Kubota:PRB49.4810,Murata:PT53.125,
PhysRevB.49.14774,Nakamura:PRB55.10549,Kawai:JPSJ68.3936,Pillay:SurfSci554.150,
Kawai:JPSJ76.034602,brand2023a,brand2024a}.
Here, we consider the transition from the
$p(2\times1)$ 
to the
$c(4\times2)$
reconstruction
at a critical temperature 
$T_{\rm crit}\approx190~\rm K$.  
Since the (001) face of single-crystalline silicon belongs to the most 
important surfaces both in technology and science, our results will 
also be relevant in this regard. 
For example, the dependence of the frozen domain structure on the 
cooling rate indicates how sufficiently homogeneous Si(001) surfaces
should be prepared.

\begin{figure}
\includegraphics[width=0.45\textwidth]{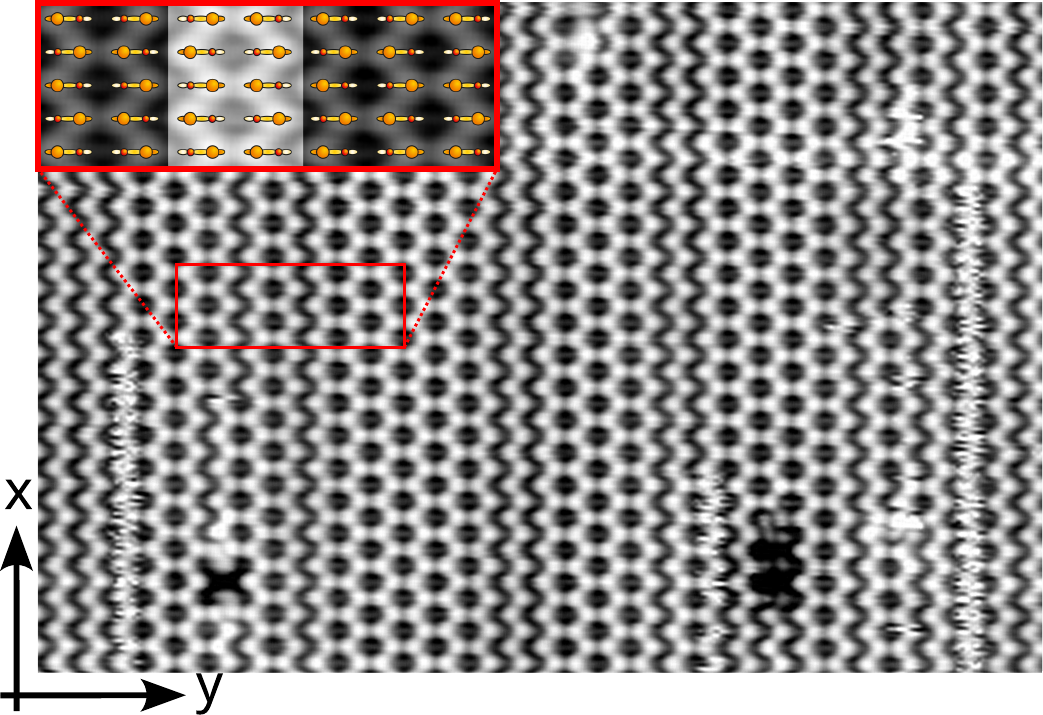}
\caption{Low temperature scanning tunneling microscope (STM) image 
of a Si(001) surface taken at 5~K with 
$U_{\rm bias} = 1.3~\rm V$ and $I_{\rm tunnel} = 1~\rm nA$. 
Field of view is $24 \times 16~\rm nm^2$. 
Areas with $c(4\times2)$ reconstruction exhibit a ``honeycomb'' pattern, 
whereas domain boundaries
can be identified by a zig-zag chain of local $p(2\times2)$ reconstruction, 
with dimer buckling and resulting domains (vertically running rows) 
indicated in the inset.
The two dark spots are frozen horizontal domain boundaries, while
frizzy vertical lines correspond to active phase boundary changes, i.e.,
mobile ''phasons''~\cite{Pennec:PRL96.026102}.}
\label{FIG:stm}
\end{figure}

\paragraph{Experimental observations} 

Let us start by presenting
experimental evidence for the formation of frozen 
domain structures on the Si(001) dimerized surface. 
The surface exhibits parallel rows of alternately buckled dimers which 
arrange in a $c(4\times2)$ reconstruction indicating the anti-phase 
correlation between neighboring dimer rows~\cite{wolkow1992direct,zhao1986atomic}. 
Fig.~\ref{FIG:stm} shows a low-temperature scanning tunneling 
microscopy (STM) image taken at 5~K after preparation of the Si(001) 
surface through flash annealing and rapid cool-down across $T_{\rm crit}$ 
to liquid nitrogen temperatures $T < 100~\rm K$. 
%
%
Further experimental details can be found 
elsewhere~\cite{Pennec:PRL96.026102}.
The STM image was taken at constant current conditions 
with positive sample bias, i.e., in Fig.~\ref{FIG:stm} filled orbitals 
of the Si atoms are displayed in bright. 
Along each vertically running row, the alternating buckling 
can be nicely identified.
The anti-phase correlation between neighboring dimers cause the 
$c(4\times2)$ reconstruction which becomes apparent as ``honeycomb'' pattern. 

During the rapid cool-down subsequent to sample preparation, the regime 
of critical slowing down~\cite{fisher1986a,tredicce2004a} is reached near $T_{\rm crit}$, where the system falls out of equilibrium,
resulting in a frozen domain structure which is apparent in Fig.~\ref{FIG:stm}.  
The domain boundaries can be identified as one-dimensional ``defects'' 
separating ordered areas with a $c(4\times2)$ reconstruction. 
As also confirmed by electron diffraction~\cite{brand2023a,brand2024a}, 
these ordered domains are extremely elongated, i.e., the frozen 
correlation length $\xi_\|$ along the dimer rows, i.e., in vertical direction, 
is much larger than the horizontal one $\xi_\perp$ across the rows. 

\new{For a rough estimate of the correlation lengths $\xi_\|$ and $\xi_\perp$,
we may count the average number of lattice sites between two defects
in Fig.~\ref{FIG:stm}.
%
%
Horizontally, this yields a correlation length $\xi_\perp$
between two and three lattice sites, whereas $\xi_\|$ is much
longer, larger than fifty lattice sites.
%
%
Thus, the anisotropy ratio $\xi_\|/\xi_\perp$ observed in Fig.~\ref{FIG:stm}
is significantly larger than the ratio of ten to one as expected from
the standard Kibble-Zurek mechanism
(as explained below and in~\bibnote[supplement]{\new{See the supplemental material}}).}
\paragraph{Anisotropic Ising model} 

In order to understand the behavior of the correlation lengths mentioned above, 
we need to model
the non-equilibrium dynamics of buckled dimers on the Si(001) 
surface which form a rectangular lattice.  
If we describe the tilt of the dimer at lattice site $i,j\in\mathbb Z$  
to the left or the right by the pseudo-spin variable 
$\sigma_{i,j}=+1$ or $\sigma_{i,j}=-1$, respectively,  
the resulting energy for the pseudo-spin configuration $\f{\sigma}$
corresponds to the anisotropic Ising 
model~\cite{Saxena:SurfSci160.618,Kubota:PRB49.4810,Murata:PT53.125,
Nakamura:PRB55.10549,Kawai:JPSJ68.3936,Pillay:SurfSci554.150,Kawai:JPSJ76.034602}
\begin{align}
\label{Ising}
E_{\fk{\sigma}}
&=& 
-J_x \sum_{i,j} \sigma_{i,j} \sigma_{i+1,j} 
-J_y \sum_{i,j} \sigma_{i,j} \sigma_{i,j+1} 
\nn
&&
-J_{\rm d} \sum_{i,j} \sigma_{i,j} [\sigma_{i+1,j+1} + \sigma_{i+1,j-1}]\,.
\end{align}
Combining microscopic considerations with experimental data, 
a strong anti-ferromagnetic coupling along dimer rows
$J_x \approx -25~{\rm meV}$
and weaker couplings across rows 
$J_y \approx 3.2~{\rm meV}$
and in diagonal direction
$J_{\rm d} \approx 2.0~{\rm meV}$
have been found~\cite{brand2023a,brand2024a}.
The latter two
can be combined into an effective transversal coupling 
$J_\perp=J_y-2J_{\rm d}\approx -0.8~{\rm meV}$.
As a result, the actual surface favors anti-ferromagnetic order 
in both directions.
For convenience however, we apply a checker-board transformation 
$\sigma_{i,j}\to(-1)^{i+j}\sigma_{i,j}$ after which both 
$J_\|=J_x$ and $J_\perp$ are positive and the transformed 
model favors ferromagnetic order.

\paragraph{Rate equations} 

We study the non-equilibrium dynamics of the Ising model~\eqref{Ising} 
via standard rate equations for the probabilities $P_{\fk{\sigma}}$ 
of the pseudo-spin configurations $\f{\sigma}$
\begin{align}
\label{rate}
\dot P_{\fk{\sigma}} 
= 
\sum_{\fk{\sigma'}}
[R_{\fk{\sigma'}\to\fk{\sigma}}P_{\fk{\sigma'}}
-
R_{\fk{\sigma}\to\fk{\sigma'}}P_{\fk{\sigma}}]
\,.
\end{align}
Neglecting correlated flips of two or more pseudo-spins (i.e., dimers),
we use single-flip transition rates 
\begin{align}
\label{Glauber}
R_{\fk{\sigma'}\to\fk{\sigma}}
=
\frac{\Gamma\exp\{-\beta E_{\rm B}\}}
{\exp\{\beta(E_{\fk{\sigma}}-E_{\fk{\sigma'}})\}+1}
\,.
\end{align}
The ``knocking'' frequency $\Gamma\approx10^{12}/{\rm s}$ and Arrhenius 
barrier height $E_{\rm B}\approx 100~{\rm meV}$ are obtained from microscopic 
considerations~\cite{PhysRevB.49.4790,Pennec:PRL96.026102,brand2024a}.
The Glauber factor in the denominator~\new{\cite{glauber1963a,sakai2013a}} can also be motivated by microscopic 
models for surface-bulk interactions, e.g., in the form of a reservoir of two-level systems~\cite{caldeira1993a} or via 
fermionic tunneling~\cite{timm2008a}.
It ensures that the rate is bounded 
$R_{\fk{\sigma'}\to\fk{\sigma}}<\Gamma\exp\{-\beta E_{\rm B}\}$
and satisfies the detailed balance condition 
$R_{\fk{\sigma'}\to\fk{\sigma}}/R_{\fk{\sigma}\to\fk{\sigma'}}=
\exp\{\new{-}\beta(E_{\fk{\sigma}}-E_{\fk{\sigma'}})\}$
which enforces evolution towards 
thermal equilibrium for constant parameters
$\Gamma$ and $\beta$ etc. 

\paragraph{1D Ising model} 

\begin{figure}
\includegraphics[width=0.45\textwidth]{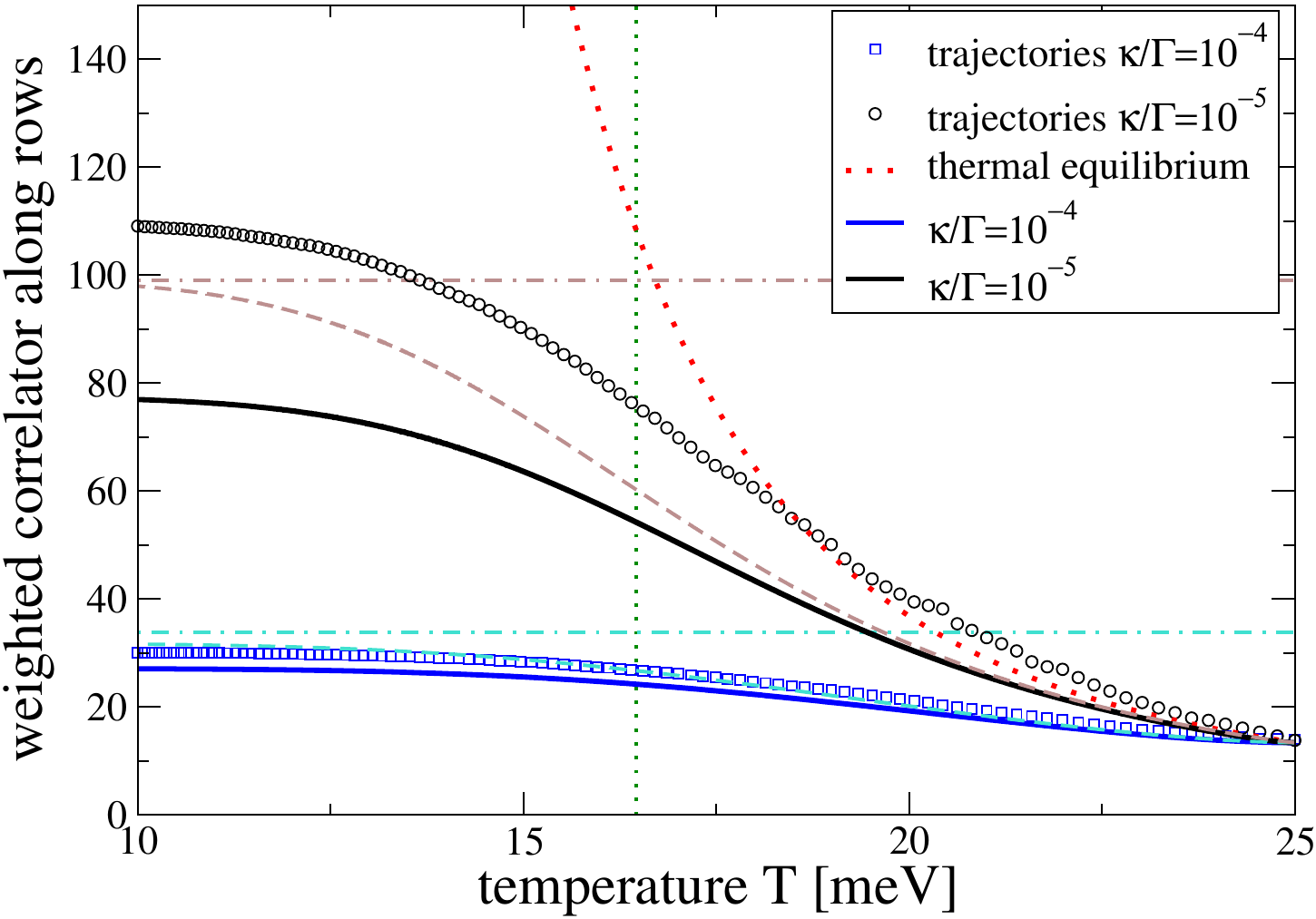}
\caption{
Freezing of the weighted sum of correlations ${\mathfrak C}$ 
for cooling quenches \new{$T(t)=T_{\rm in} e^{-\kappa t}$}.
The black solid curve represents the \new{numeric} solution of Eq.~\eqref{weighted}
for the 1D Ising model~\eqref{1D-Ising} 
for \new{$\kappa=10^7~\rm s^{-1}$} 
and the \new{solid} blue curve for
\new{$\kappa=10^8~\rm s^{-1}$}, 
while the symbols correspond to full numerical simulations 
of the 2D Ising model~\eqref{Ising} on a $1600\times 200$
lattice averaged over $100$ trajectories. 
Correlations between rows remain \new{weak}
for these parameters, \new{see below.}
The dotted red curve depicts the \new{thermal} equilibrium value in 1D and
\new{the dashed curves and dash-dotted horizontal lines in lighter colors represent the analytic solutions of the equation
$\partial_{\mathfrak T} \mathfrak C = \lambda \mathfrak T \mathfrak C + 1$
and the approximations
${\mathfrak C}_{\rm freeze}\approx\sqrt{\pi/(2\lambda)}$
to the frozen values discussed in the text~\bibnotemark[supplement].}
The vertical dotted line depicts the critical temperature in 2D. 
}
\label{FIG:1D}
\end{figure}

Now we are in the position to study the non-equilibrium dynamics of the 
Ising model~\eqref{Ising} during a cooling quench \new{$\beta\to\beta(t)$}.
As the first step, we consider a very rapid cooling rate such that the 
system basically has no time for an exchange between the dimer rows, i.e., 
in the weakly coupled direction. 
Then, as also confirmed by numerical simulations (see Fig.~\ref{FIG:1D}), 
we may 
consider the limiting case $J_y\to0$ and $J_{\rm d}\to0$ of the 2D Ising 
model~\eqref{Ising} such that
each row $j$ separately forms a 1D Ising model 
with $J=J_x=J_\|$
\begin{align}
\label{1D-Ising}
E_{\fk{\sigma}}^{\rm 1D}
= 
-J \sum_{i} \sigma_{i} \sigma_{i+1} 
\,,
\end{align}
where after the checker-board transformation $J>0$. 
Assuming translational invariance, we may derive an exact evolution equation 
for the correlator $c_a=\langle\sigma_{i}\sigma_{i+a}\rangle$ depending on 
distance $a$.
With the dimensionless conformal time coordinate
$d{\mathfrak T}/dt=\Gamma e^{-\beta(t)E_{\rm B}}$ 
we find (with the boundary condition $c_{a=0}=1$)
\bibnotemark[supplement]
\begin{align}
\label{conformal}
\new{
\partial_{\mathfrak T}c_a=
-2c_a
+(c_{a+1}+c_{a-1})
\tanh[2\beta({\mathfrak T})J]
\,. 
}
\end{align}
Setting the left-hand side to zero
yields the well-known equilibrium solution $c_a=[\tanh(\beta J)]^{|a|}$. 
The 1D Ising model~\eqref{1D-Ising} does not have a critical 
point at finite temperature $T_{\rm crit}>0$, instead the analogue of 
a critical point occurs at zero temperature $T_{\rm crit}=0$ where the 
correlation length $\xi$ diverges as $\xi\sim e^{2\beta J}$~\cite{baxter1989}. 

In order to understand the non-equilibrium dynamics governed by 
Eq.~\eqref{conformal}, 
let us consider the continuum limit, where $c_{a+1}+c_{a-1}-2c_a$
becomes the second spatial derivative such that we obtain a 
diffusion-dissipation equation 
$\partial_{\mathfrak T}c={\mathfrak D}\partial_x^2c-\gamma c$.
For large temperatures, the diffusion coefficient is small 
${\mathfrak D}\propto\tanh(2\beta J)\approx2\beta J\ll1$ 
and the damping term $\gamma\approx2$ dominates. 
For small temperatures, the damping rate $\gamma=2-2\tanh(2\beta J)$ 
is suppressed as $4e^{-4\beta J}$ and the diffusion term 
${\mathfrak D}\propto\tanh(2\beta J)\approx1$ 
dominates.
Thus, we may introduce a response or relaxation time 
${\mathfrak T}_{\rm relax}$ from the inverse damping rate $1/\gamma$ 
which then scales as 
${\mathfrak T}_{\rm relax}\sim e^{4\beta J}$, i.e., 
${\mathfrak T}_{\rm relax}\sim\xi^2$. 
Note, however, that the diffusion coefficient stays finite even for 
${\mathfrak T}_{\rm relax}\to\infty$, i.e., diffusion is still possible.

\paragraph{Freezing in 1D} 

Since analyzing the non-equilibrium dynamics by means of analytic solutions of 
Eq.~\eqref{conformal} is still quite involved, let us consider the weighted 
sum of correlations ${\mathfrak C}=\sum_{a=1}^\infty a c_a$ which obeys 
%
\begin{align}
\label{weighted}
\partial_{\mathfrak T}{\mathfrak C}
=
-2[1-\tanh(2\beta J)]
{\mathfrak C}
+\tanh(2\beta J)
\,.
\end{align}
In order to provide an explicit example and to study the analogue of critical
slowing down, let us assume the simple cooling protocol
\new{$T(t)=T_{\rm in}e^{-\kappa t}$ or $\beta(t)=\beta_{\rm in}e^{\kappa t}$
starting at an initial temperature $T_{\rm in}$
and cooling down to zero temperature at $t\to\infty$.
This protocol can be motivated by assuming that the system is cooled down
via ordinary thermal conduction to a cold reservoir, for example.
Then the conformal time is given by the exponential integral
${\mathfrak T}(t)={\rm Ei}(-e^{\kappa t}E_{\rm B}/T_{\rm in})\Gamma/\kappa$
and the infinite interval of laboratory time $t\in(0,\infty)$ is mapped
to a finite interval of conformal time
${\mathfrak T}\in({\mathfrak T}_{\rm in},0)$.}

%

\new{
Incidentally, for our values with $E_{\rm B}\approx4 J_x$ and assuming
$E_{\rm B}\gg T_{\rm in}$, we may simplify
Eq.~\eqref{weighted} even further.
In this limit, we may approximate
the exponential integral by its asymptotic behavior
${\rm Ei}(-z)\approx-e^{-z}/z$ and
Eq.~\eqref{weighted} becomes
$\partial_{\mathfrak T}{\mathfrak C}\approx
\lambda{\mathfrak T}{\mathfrak C}+1$ with
$\lambda=4\kappa E_{\rm B}/(T_{\rm in}\Gamma)$.}
%
%
%
%
%
%
The solution to this equation can be given in terms of the error function~\bibnotemark[supplement], 
but we may understand its behavior by means of general arguments.
%
\new{Since we have $|{\mathfrak T}_{\rm in}|\gg1$ for our parameters,
let us start at ${\mathfrak T}_{\rm in}\to-\infty$.
First, ${\mathfrak C}$ approaches its instantaneous
equilibrium value ${\mathfrak C}_{\rm eq}=1/|\lambda{\mathfrak T}|$.
}
%
%
%
%
However, once the response time 
\new{${\mathfrak T}_{\rm relax}\sim1/|\lambda{\mathfrak T}$|}
%
%
becomes too short
${\mathfrak T}_{\rm relax}\sim|{\mathfrak T}|$, the system cannot 
equilibrate anymore and thus the value of ${\mathfrak C}$ freezes in at 
\new{${\mathfrak T}_{\rm freeze}\sim1/\sqrt{\lambda}$}
%
%
to its final value
\new{${\mathfrak C}_{\rm freeze}\approx\sqrt{\pi/(2\lambda)}$.}
%
%
%
%
Furthermore, as $\mathfrak C$ scales with the square of the correlation
length $\xi$, we obtain 
\new{$\xi_{\rm freeze}\sim\lambda^{-1/4}\sim\kappa^{-1/4}$,}
%
%
which is the analogue to the Kibble-Zurek scaling for this case.
\new{E.g., for the value of $\kappa=10^7~\rm s^{-1}$ used in the solid black curve in
Fig.~\ref{FIG:1D}, we have $\lambda=\ord(10^{-4})$ and thus
$\xi_{\rm freeze}=\ord(10)$.
Comparison with Fig.~\ref{FIG:stm} indicates that the cooling rate
in the experiment must have been smaller.} 

\paragraph{2D Ising model} 

As the next step, let us study
slower cooling rates, where we start to see two-dimensional behavior.
For simplicity, we consider the two-dimensional Ising model in terms 
of the effective transversal coupling strength $J_\perp$ introduced above 
\begin{align}
\label{2D-Ising}
E_{\fk{\sigma}}^{\rm 2D}
= 
-J_\| \sum_{i,j} \sigma_{i,j} \sigma_{i+1,j} 
-J_\perp\sum_{i,j} \sigma_{i,j} \sigma_{i,j+1} 
\,.
\end{align}
The equilibrium properties of this model can be obtained by Onsager 
theory~\cite{Onsager:PR65.117}.
Its critical temperature $T_{\rm crit}=1/(k_{\rm B}\beta_{\rm crit})$ 
is determined by the relation
$\sinh(2\beta_{\rm crit}|J_\||)\sinh(2\beta_{\rm crit}|J_\perp|)=1$. 
Thus, in the limit of strong anisotropy $J_\|\gg J_\perp$, we obtain the 
hierarchy of scales $J_\| \gg\beta_{\rm crit}^{-1}\gg J_\perp$. 
Approaching the critical point from above, 
the correlation lengths $\xi_\|$ and $\xi_\perp$ in $x$- and $y$-direction 
(i.e., along the dimer rows and perpendicular to them) 
both obey the scaling 
$|T-T_{\rm crit}|^{-\nu}$ with the critical exponent $\nu=1$,
though with different pre-factors~\cite{mccoy1973,HobrechtHucht:SciPostPhys7}.
In particular, their ratio stays constant and is given by
$\xi_\perp^{\rm eq}/\xi_\|^{\rm eq}=
\sinh(2\beta_{\rm crit}J_\perp)\approx2\beta_{\rm crit}J_\perp\approx0.1$. 
\new{Thus, in the standard Kibble-Zurek picture~\bibnotemark[supplement] where both correlations
freeze at time ${\mathfrak T}_{\rm freeze}$, one would expect the
same ratio for their frozen values.}

Now considering the correlator 
$c_{a,b}=\langle\sigma_{i,j}\sigma_{i+a,j+b}\rangle$
and its evolution equation analogous to Eq.~\eqref{conformal},
we find that the terms on the right-hand side containing $J_\perp$
do also involve four-point correlators such as 
$\langle
\sigma_{i,j}\sigma_{k-1,\ell}\sigma_{k+1,\ell}\sigma_{k,\ell\pm1}
\rangle$.
To close this set of equations approximately, we may apply a perturbative 
expansion scheme and neglect terms of order $J_\perp^2$~\bibnotemark[supplement].
Then, since the four-point correlators have already \new{a} small pre-factor
$\sim J_\perp$ out front, we may approximate them by their zeroth order 
$\langle
\sigma_{i,j}\sigma_{k-1,\ell}\sigma_{k+1,\ell}\sigma_{k,\ell\pm1}
\rangle
\approx 
\langle\sigma_{i,j}\sigma_{k,\ell\pm1}\rangle
\langle\sigma_{k-1,\ell}\sigma_{k+1,\ell}\rangle$
where the short-range correlations 
$\langle\sigma_{k-1,\ell}\sigma_{k+1,\ell}\rangle$
within a row $\ell$ can be approximated by their equilibrium value
$\langle\sigma_{k-1,\ell}\sigma_{k+1,\ell}\rangle\approx
\tanh^2(\beta J_{\|})$ in 1D (as discussed above). 
After that, we obtain an approximate diffusion-dissipation equation in 2D 
\begin{align}
\label{2D-correlation}
\partial_{\mathfrak T}
c_{a,b}
&=&
-2c_{a,b}
+(c_{a+1,b}+c_{a-1,b})
\tanh(2\beta J_\|)
\nn
&&
+
\beta J_\perp^{\rm eff}(c_{a,b+1}+c_{a,b-1})
+
\ord(J_\perp^2) 
\,,
\end{align}
but with a strongly reduced transversal coupling strength 
$J_\perp^{\rm eff}=2J_\perp/\cosh(2\beta J_\|)$.  
As an intuitive picture, 
the pre-existent strong in-row correlation requires many spin flips 
for cross-row equilibration and thus renders this process very hard.
In the continuum limit, the ratio between the effective diffusion
coupling strengths ${\mathfrak D}_{\|} \propto \tanh(2\beta J_\|)$ and 
${\mathfrak D}_{\perp} \propto \beta J_\perp^{\rm eff}$ in the two 
directions in Eq.~\eqref{2D-correlation} determines the ratio of the 
equilibrium correlation lengths which is consistent with the results above
$\xi_\perp^{\rm eq}/\xi_\|^{\rm eq}
=
\sqrt{{\mathfrak D}_{\perp}/{\mathfrak D}_{\|}} = \sqrt{2\beta J_\perp/\sinh(2\beta J_{\|})}
\approx
2\beta_{\rm crit}J_\perp$
close to the critical point.  

In the non-equilibrium case, Eq.~\eqref{2D-correlation} allows us to 
understand the crossover from 1D to 2D. 
Inserting our values, we find $\beta_{\rm crit}J_\perp^{\rm eff}|_{\rm crit}\approx 0.01$, 
which explains why a time interval of
\new{$|{\mathfrak T}_{\rm in}|\approx 37.8$}
(\new{solid} blue curve in Fig.~\ref{FIG:1D}) is too short to generate correlations between the rows.
Only for slower sweep rates, such as
\new{$|{\mathfrak T}_{\rm in}|\approx 378$}
(solid black curve in Fig.~\ref{FIG:1D}),
these correlations start to build up, 
even though they are still very weak.
\new{Actually, by employing time-dependent perturbation theory in
$J_\perp$ via inserting the time-dependent (non-equilibrium)
correlator $\langle\sigma_{k-1,\ell}\sigma_{k+1,\ell}\rangle\approx c_{a=2}(t)$
within a row $\ell$ obtained from the exact 1D solution discussed
in Eq.~\eqref{conformal}
into the evolution equation for
$\partial_{\mathfrak T}c_{a,b}$, we can describe the build-up of
correlations between rows reasonably well for fast and intermediate
sweep rates up to $\kappa=10^6~\rm s^{-1}$~\bibnotemark[supplement].}

\paragraph{Numerical simulations} 

Finally, let us compare the analytical approximation schemes discussed above 
to a full numerical simulation of Eqns.~(\ref{Ising}-\ref{Glauber})
with time-dependent $\beta(t)$~\bibnotemark[supplement]. 
Due to the exponential dimensionality of~\eqref{rate} for an $N_x\times N_y$ 
spin lattice, we calculate trajectory solutions.
For a given configuration $\f{\sigma}$, 
we propagate time by the stochastic waiting 
time $\tau_{\fk{\sigma}}$ found by numerically solving
$\ln(1-r)=-\sum_{\fk{\sigma'}} \int_{t}^{t+\tau_{\fk{\sigma}}}
R_{\fk{\sigma}\to\fk{\sigma'}}(t') dt'$,
with uniformly distributed random number $r\in[0,1]$, 
and perform a jump to a different state with the conditional 
probability~\cite{bulnes1998a,Binder2019}
given by 
$P_{\fk{\sigma}\to\fk{\sigma'}} = R_{\fk{\sigma}\to\fk{\sigma'}}/
[\sum_{\fk{\sigma''}\neq \fk{\sigma}} R_{\fk{\sigma}\to\fk{\sigma''}}]$ 
at time $t+\tau_{\fk{\sigma}}$.
In the selection of jumps, we use~\cite{kratzer2009a}
that the $N_x N_y$ different single-spin flip processes can be 
grouped into 45 classes with identical energy differences entering 
the rates~\eqref{Glauber}.
Eventually, denoting the fast Fourier transformed spin lattice by 
$\tilde{\sigma}_{k_x k_y}$, the correlation lengths $\xi_{\|}$ and $\xi_{\perp}$
are then given by the inverse widths of the one-dimensional 
lattices $\sum_{k_y}|\tilde{\sigma}_{k_x k_y}|^2$ and 
$\sum_{k_x}|\tilde{\sigma}_{k_x k_y}|^2$, respectively. 
Averaging over multiple trajectories (and the resulting 
$|\tilde{\sigma}_{k_x k_y}|^2$) can be used to improve the 
statistics.

\begin{figure}
\hspace{0.5cm}\includegraphics[width=0.45\textwidth]{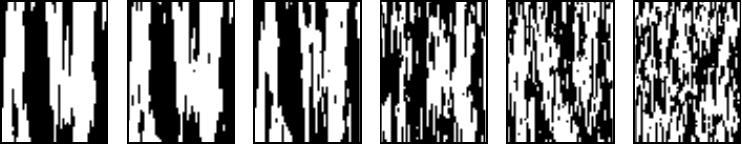}\\
\includegraphics[width=0.45\textwidth]{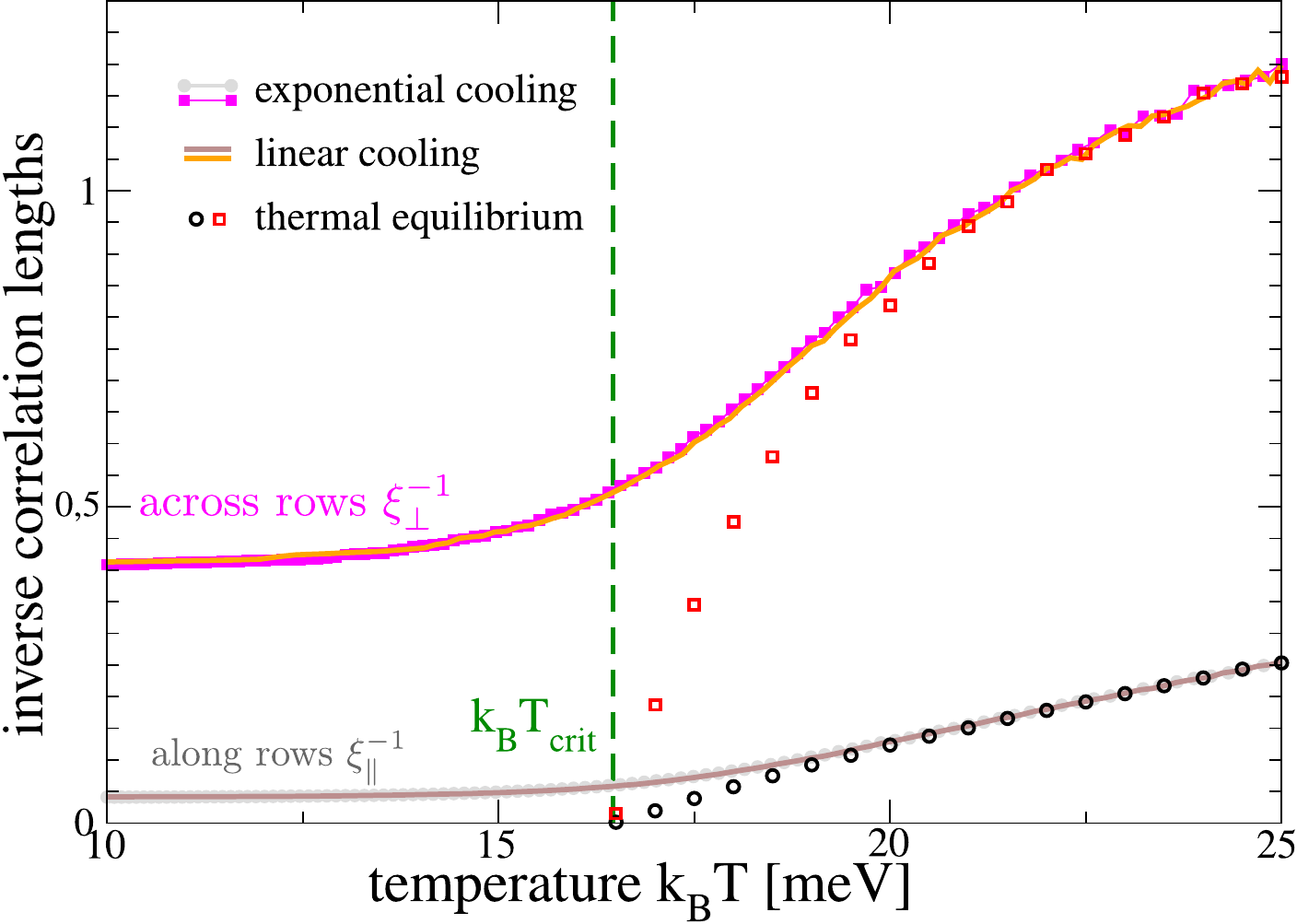}
\caption{Plot of inverse correlation lengths versus temperature
(or, equivalently, time) for \new{exponential $T(t)=T_{\rm in} e^{-\kappa t}$ (connected symbols) and linear $T(t)=T_{\rm in}-\eta t$ (solid curves)
cooling sweeps.
Choosing $\eta=\kappa T_{\rm crit}$ ensures that the cooling rates
at the critical points coincide -- as a result, the curves for the
two protocols lie almost on top of each other.
The value $\eta=\kappa T_{\rm crit}=174\times10^6~\rm K/s$ used here
corresponds to $\Gamma/\kappa \approx 10^6$ in Fig.~\ref{FIG:kibblezurek}.
The results are obtained for a $16000\times 2000$ lattice, averaged
over $100$ trajectories.}
%
The correlation length \new{$\xi_\perp$} in weakly-coupled direction
departs earlier than the other \new{$\xi_\|$}
from the equilibrium solutions (red squares and black circles).
On top, we added
\new{lattice portion snap-shots of $67\times 50$ pseudo-spins showing the (exponential protocol) 
time evolution of an example
configuration at the respective temperatures as an illustration.}
}
\label{FIG:clengthvstemp}
\end{figure}

In Fig.~\ref{FIG:clengthvstemp}, we contrast the time-dependent 
averaged correlation lengths (connected symbols and solid curves)
with equilibrium versions (symbols) for \new{exponential and linear cooling protocols (starting after an equilibration phase).}
%
%
Already at temperatures above $T_{\rm crit}$, the correlation 
lengths depart from their equilibrium limits, but furthermore we 
see that this happens earlier for the weakly coupled direction.

\begin{figure}
\includegraphics[width=0.45\textwidth]{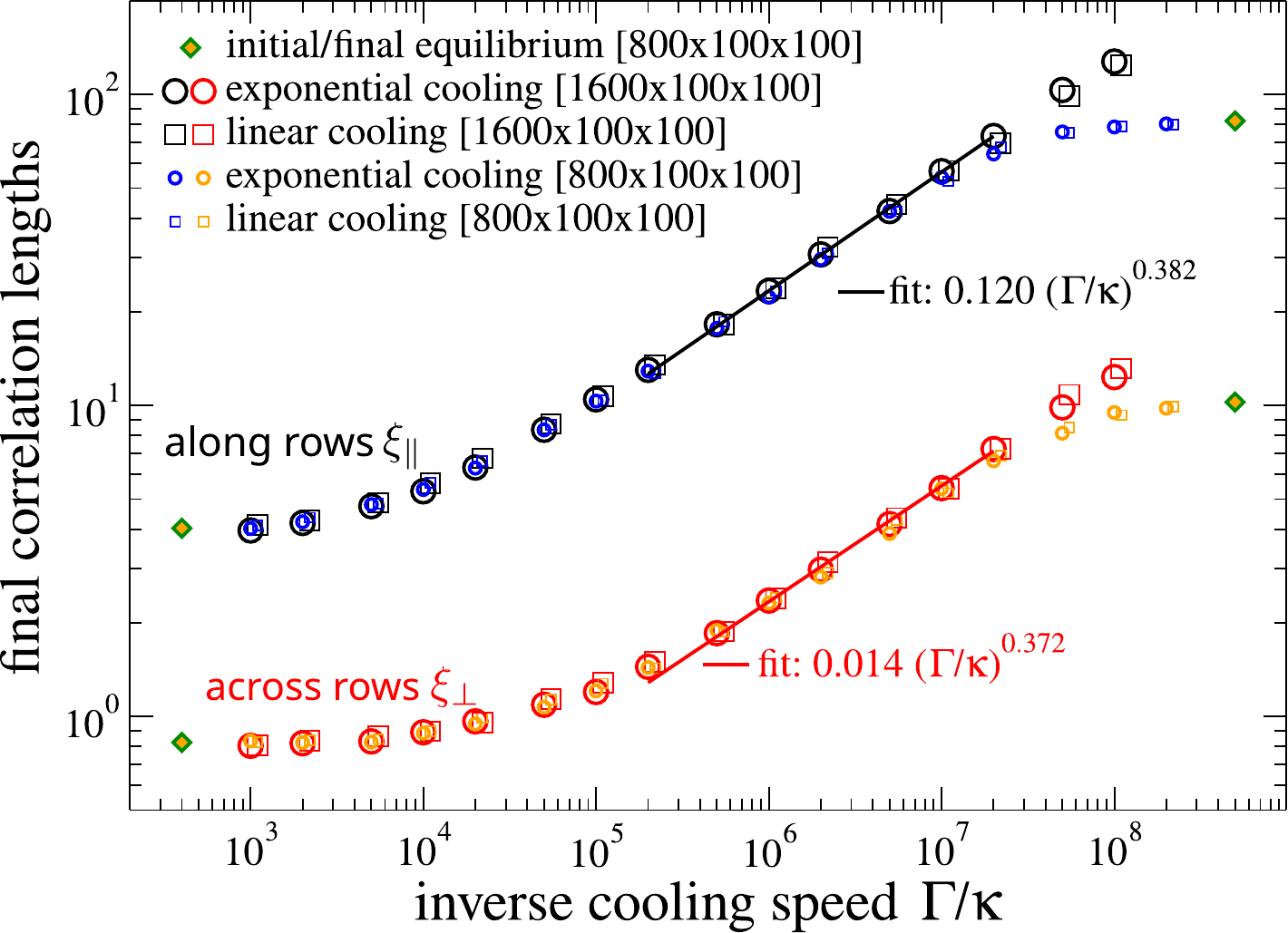}
\caption{
Plot of the final (frozen) correlation lengths for different
lattice sizes and averaged over $100$ trajectories 
$[N_x \times N_y \times N_{\rm trj}]$
\new{versus the inverse critical cooling speed
$\Gamma/\kappa$
for exponential ($T(t) = T_{\rm in} e^{-\kappa t}$, circle symbols)
and linear
($T(t) = T_{\rm in} - \eta t$ with $\eta=\kappa T_{\rm crit}$
as in Fig.~\ref{FIG:clengthvstemp}, square symbols)
protocols, all cooling the system down from
$k_{\rm B}T_{\rm in}=25~\rm meV$ to $k_{\rm B}T_{\rm fi}=10~\rm meV$.}
For too fast protocols (left), the system can never follow, 
but Kibble-Zurek scaling is recovered for intermediate protocol times. 
Finite-size effects are visible for slow protocols \new{(small vs. large symbols)}.
}
\label{FIG:kibblezurek}
\end{figure}

The final (i.e., frozen) correlation lengths are depicted 
in Fig.~\ref{FIG:kibblezurek} as a function of 
\new{the inverse cooling rate $\Gamma/\kappa$ for exponential
(circle symbols) and linear 
(square symbols) cooling protocols.
}
For extremely fast sweeps \new{$\Gamma/\kappa<10^3$}, the system cannot
follow and basically remains at the initial equilibrium values.
However, already for \new{$\Gamma/\kappa=\ord(10^4)$}, we start observing
a growth of $\xi_\|$ while $\xi_\perp$ is still negligible, i.e., 
effectively 1D behavior \new{(cf.~the blue curve in Fig.~\ref{FIG:1D})}.
For intermediate-speed sweeps, such as \new{$\Gamma/\kappa=\ord(10^6)$},
we find that both final correlation lengths follow a universal power-law 
increase, consistent with the Kibble-Zurek exponent $\nu/(1+z\nu)\approx 1/3$,  
see the fitted regions in Fig.~\ref{FIG:kibblezurek}.
For very slow sweeps, we find that finite-size effects start to play a role. 
\new{Experimentally, such finite-size effects may origin from the
omnipresent steps of a real Si(001) surface~\cite{tromp1993a},
which ultimately limit the correlation lengths at low cooling rates.
As we are interested in higher cooling rates in order to observe the
frozen domain structure, this can be safely avoided for samples with low
miscut -- we have used a wafer with $<\pm 0.25^{\circ}$ precision in Fig.~\ref{FIG:stm}.}

\paragraph{Conclusions} 

As a model for the buckling dynamics of dimers on Si(001) surfaces, 
we consider the anisotropic Ising model in 2D and study the freezing 
of spatial correlations (which determine the final domain structure) 
when cooling through the critical point
\new{via $\beta(t)=\beta_{\rm in}e^{\kappa t}$.}
Depending on the cooling rate \new{$\kappa$},
we find a crossover from effectively
1D behavior for rapid cooling to 2D behavior for slower cooling rates. 
Similar crossovers can also be observed for symmetry-breaking fields~\cite{suzuki2024a}.
Exact analytic solution of the 1D case
yields a scaling of the frozen 
correlation lengths \new{$\xi_{\rm freeze}\propto\kappa^{-1/4}$.}
For the 2D case 
at intermediate cooling rates \new{$\kappa$},
we numerically found a scaling with
\new{$\kappa^{-\nu/(1+z\nu)}\approx\kappa^{-1/3}$}
which is consistent with the Kibble-Zurek scaling in 2D.
\new{When comparing our results (e.g., that the cooling protocol can affect the anisotropy) with the experimental data in Fig.~\ref{FIG:stm},
we find that they match qualitatively but not quantitatively.
Note that the exact cooling protocol for the sample in Fig.~\ref{FIG:stm}
is not precisely known.
Whether the mismatch is caused by annealing after traversing the critical
point (predominantly within the rows) or by the fact that parameters,
such as the barrier height $E_{\rm B}$ and shape in Eq.~\eqref{Glauber},
effectively deviate from the values we used should be investigated in
future experiments by systematically varying the cooling rate.
In this way, one can also turn the argument around:
While the parameters in the Ising model~\eqref{2D-Ising}
can be obtained from equilibrium measurements~\cite{brand2023a,brand2024a},
these non-equilibrium scenarios facilitate experimental access to other parameters such as the knocking frequency $\Gamma$ or the barrier height
$E_{\rm B}$.
Our findings encourage further experimental investigations of the
frozen domain structure for varying cooling rates
in this anisotropic 2D Ising system in order to provide a
well-controlled experimental approach to Kibble-Zurek dynamics
in condensed matter.

%

}

%
%

\acknowledgments

\paragraph{Acknowledgments} 

Funded by the Deutsche Forschungsgemeinschaft 
(DFG, German Research Foundation) through 
the Collaborative Research Center
SFB~1242 ``Nonequilibrium dynamics of condensed matter
in the time domain'' (Project-ID 278162697).

\bibliographystyle{unsrt}
\bibliography{references.bib}

\clearpage
{\Large Supplemental material}

\section{Appendix: Kibble-Zurek scaling}

Let us briefly recapitulate the main arguments leading to the standard 
Kibble-Zurek scaling.
We consider a symmetry-breaking second-order phase transition at the critical 
temperature $T_{\rm crit}$.
Approaching 
the critical point $T_{\rm crit}$, the equilibrium correlation length 
$\xi^{\rm eq}$ obeys the universal scaling behavior
\begin{align}
\label{xi}
\xi^{\rm eq} \sim \left|\frac{T-T_{\rm crit}}{T_{\rm crit}}\right|^{-\nu}\equiv
|\tau|^{-\nu}
\end{align}
with the universal critical exponent $\nu$ and the dimensionless reduced 
temperature $\tau$. 
Similarly, the response or relaxation time $t_{\rm relax}^{\rm eq}$
(in equilibrium) scales as 
\begin{align}
\label{tau}
t_{\rm relax}^{\rm eq}
\sim
|\tau|^{-z\nu}
\sim(\xi^{\rm eq})^z
%
\end{align}
with the dynamical critical exponent $z$. 
The divergence of $t_{\rm relax}^{\rm eq}$ at the critical point is the hallmark 
of critical slowing down~\cite{fisher1986a,tredicce2004a}. 

Now the idea is to infer non-equilibrium properties from these equilibrium 
values. 
Let us assume a linear time-dependence for the cooling protocol 
$\tau(t)=-\eta t$ 
with the cooling rate $\eta$ such that, starting at the time $t_{\rm in}<0$, 
the critical point is reached at $t=0$.
Then we may estimate the freezing time via 
$|t_{\rm freeze}|=t_{\rm relax}^{\rm eq}(\tau_{\rm freeze})$
where $\tau_{\rm freeze}=\tau(t_{\rm freeze})$,
after which the system has no time to equilibrate any more. 
Insertion into Eq.~\eqref{tau} yields 
$|t_{\rm freeze}|\sim\eta^{-z\nu/(z\nu+1)}$
and thus we obtain the frozen correlation length from Eq.~\eqref{xi}
\begin{align}
\label{scaling}
\xi_{\rm freeze}=\xi^{\rm eq}(\tau_{\rm freeze})\sim\eta^{-\nu/(z\nu+1)}
\,,
\end{align}
which is referred to as Kibble-Zurek scaling~\cite{dutta2010a,ulm2013a,qiu2020a}.

\section{Appendix: 1D Ising model} 

A closer look at the Glauber-Arrhenius rates~(3)
shows that for 
the energy differences 
arising for a single-spin flip at site $i$ in the 1D Ising model ($\sigma_i'=-\sigma_i$)
\begin{align}
E_{\fk{\sigma}}-E_{\fk{\sigma}'} = - 2 J \sigma_i (\sigma_{i-1}+\sigma_{i+1})\,,
\end{align}
the rates can be represented by a quadratic function of the spins near that 
site~\cite{glauber1963a,sakai2013a} with 
\begin{align}
\frac{1}{e^{\beta(E_{\fk{\sigma}'}-E_{\fk{\sigma}})}+1} = 
\frac{1}{2} - \frac{\tanh(2\beta J)}{4} \sigma_i (\sigma_{i-1}+\sigma_{i+1})
\,.
\end{align}
For the expectation value of a spin at site $i$ this implies via Eq.~(2)
\begin{align}\label{mean-field}
\partial_t \langle \sigma_i \rangle &= \sum\limits_{\fk{\sigma}} \sigma_i \dot P_{\fk{\sigma}} = \sum\limits_{\fk{\sigma},\fk{\sigma}'} (\sigma_i'-\sigma_i) R_{\fk{\sigma}\to\fk{\sigma}'} P_{\fk{\sigma}}\\
&= -2 \sum\limits_{\fk{\sigma}} \sigma_i R_{\fk{\sigma} \to F_i \fk{\sigma}} P_{\fk{\sigma}}\nn
&= - \Gamma e^{-\beta E_{\rm B}} \times\nn
&\qquad\times \sum\limits_{\fk{\sigma}}  \left[\sigma_i -\frac{\tanh(2\beta J)}{2}(\sigma_{i-1}+\sigma_{i+1})\right]P_{\fk{\sigma}}\nn
&=
\Gamma e^{-\beta E_{\rm B}}
\left[
\tanh(2\beta J)
\frac{\langle\sigma_{i+1}\rangle+\langle\sigma_{i-1}\rangle}{2}
-\langle\sigma_i\rangle
\right]\,,\nonumber
\end{align}
where $F_i$ denotes the flip operator for the $i$th spin and we used furthermore $\sigma_i^2=1$.
With this, we have obtained exact evolution equations for the non-equilibrium expectation values.

%
%
Similarly, we can write the two-point functions as
$\partial_t \langle \sigma_i \sigma_j \rangle = -2 \sum_{\fk{\sigma}} \sigma_i \sigma_j (R_{\fk{\sigma} \to F_i \fk{\sigma}} + R_{\fk{\sigma}\to F_j \fk{\sigma}}) P_{\fk{\sigma}}$, 
which eventually leads to
\begin{align}
\label{two-point}
\partial_t\langle\sigma_i\sigma_j\rangle
=
\Gamma e^{-\beta E_{\rm B}}
\left[ 
-2\langle\sigma_i\sigma_j\rangle
+\tanh(2\beta J)\times 
\phantom{\frac12}
\right. 
\nn
\left.
\times 
\frac{
\langle\sigma_{i+1}\sigma_{j}\rangle+
\langle\sigma_{i-1}\sigma_{j}\rangle+ 
\langle\sigma_{i}\sigma_{j+1}\rangle+
\langle\sigma_{i}\sigma_{j-1}\rangle
}{2}
\right] 
\,.
\end{align}
If we start in the symmetric phase $\langle\sigma_i\rangle=0$, 
the mean field $\langle\sigma_i\rangle$ in Eq.~\eqref{mean-field}
stays zero as expected. 
However, due to 
$\langle\sigma_{i}\sigma_{i}\rangle=\langle\sigma_{i}^2\rangle=1$, 
the evolution equation~\eqref{two-point} for the correlations 
contains a source term and thus correlations are generated even 
if they are absent initially. 

Taking $j=i+a$ and assuming translational invariance, we obtain from~\eqref{two-point}
an equation for the correlator
\begin{align}
\partial_t c_a = \Gamma e^{-\beta E_{\rm B}} \left[-2 c_a + \tanh(2\beta J) (c_{a-1} + c_{a+1})\right]\,,
\end{align}
which upon transformation to the conformal time coordinate becomes Eq.~(5) 
in the main text.

\section{Appendix: Freezing in 1D} 

\subsection{Exponential cooling sweep $T(t) = T_{\rm in} e^{-\kappa t}$}

\begin{figure}
\includegraphics[width=0.5\textwidth]{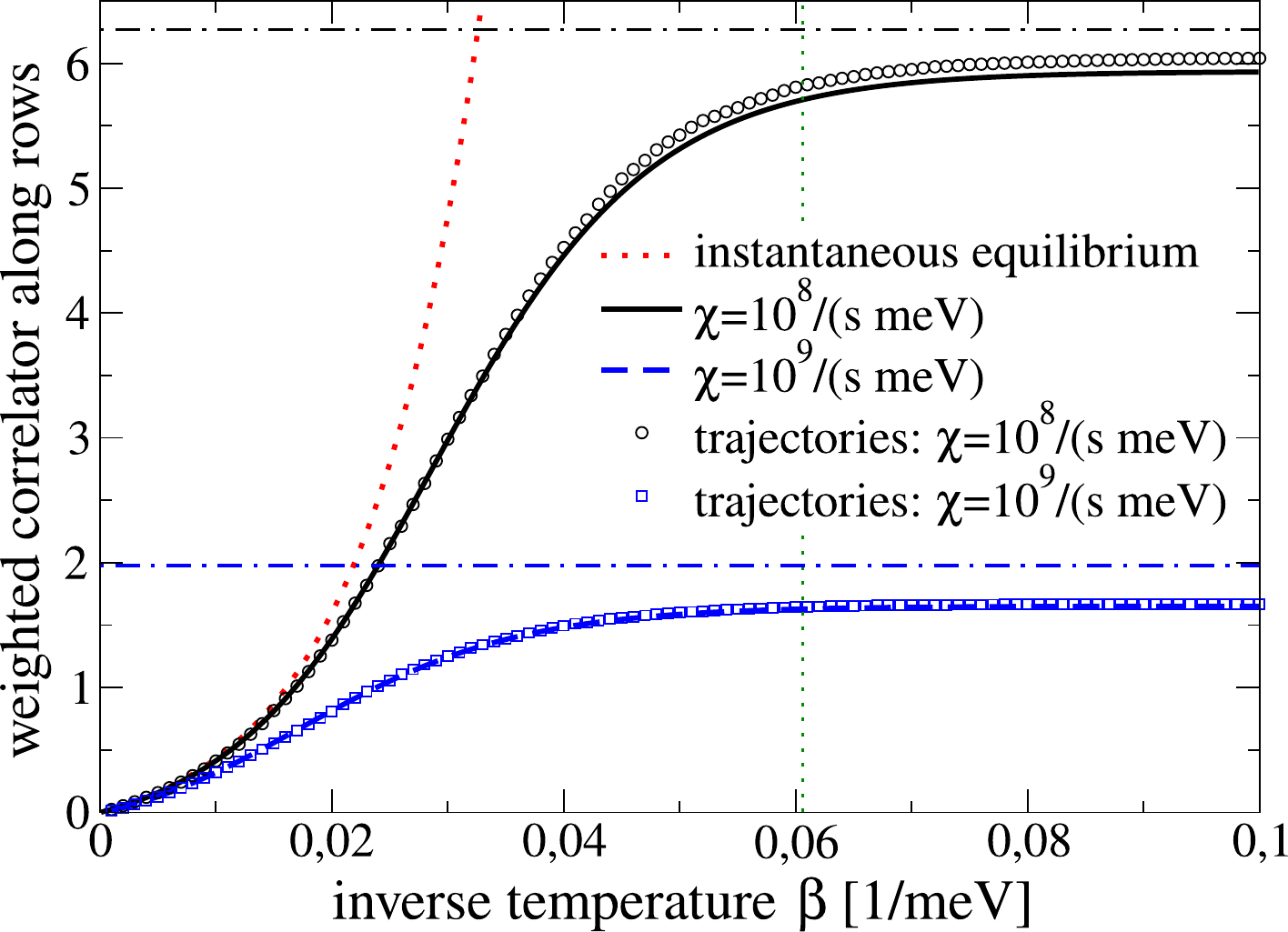}
\caption{
Freezing of the weighted sum of correlations ${\mathfrak C}$ 
for cooling quenches \new{$\beta(t)=\chi t$}.
The black solid curve represents the analytic solution of Eq.~(6)
for the 1D Ising model~(4)
for \new{$\chi=10^8~\rm s^{-1} meV^{-1}$} 
and the dashed blue curve for 
\new{$\chi=10^9~\rm s^{-1} meV^{-1}$} 
while the symbols correspond to full numerical simulations 
of the 2D Ising model~(1)
on a $1600\times 200$ lattice averaged over $100$ trajectories.
Correlations between rows remain negligible for these parameters (not shown).
The dotted red curve depicts the \new{thermal} equilibrium value in 1D and the horizontal 
dashed-dotted lines represent the approximation 
${\mathfrak C}_{\rm freeze}\approx\sqrt{\pi|{\mathfrak T}_{\rm in}|/8}$
for the two cases with 
${\mathfrak T}_{\rm in}=-10$ 
and ${\mathfrak T}_{\rm in}=-100$, 
respectively. 
The vertical dotted line depicts the critical temperature in 2D. 
}
\label{FIG:1Dold}
\end{figure}

\new{
The conformal time mapping obeying the differential equation $d\mathfrak T/dt = \Gamma e^{-\beta(t) E_{\rm B}}$ is given by 
\begin{align}\label{EQ:conf_time}
{\mathfrak T}(t)={\rm Ei}\left(-e^{\kappa t}\frac{E_{\rm B}}{T_{\rm in}}\right)\frac{\Gamma}{\kappa}\,,
\end{align}
where ${\rm Ei}(x)=-\int_{-x}^\infty \frac{e^{-t}}{t} dt$ is the exponential integral. 
For large negative arguments, it has the asymptotic expansion ${\rm Ei}(-x) \approx -e^{-x}/x$, such that 
expanding the exponential integral for $E_{\rm B} \gg T_{\rm in}$, we obtain (but see also Eqns.~\eqref{EQ:exp_approx} and~\eqref{EQ:plog_approx} for further improvements)
\begin{align}
{\mathfrak T}(t) &\approx -\frac{\Gamma T_{\rm in}}{\kappa E_{\rm B}} \exp\left\{-e^{\kappa t} \frac{E_{\rm B}}{T_{\rm in}}-\kappa t\right\}\nn
&\approx -\frac{\Gamma T_{\rm in}}{\kappa E_{\rm B}} \exp\left\{-e^{\kappa t} \frac{E_{\rm B}}{T_{\rm in}}\right\} 
\equiv {\mathfrak T}_{\rm in} e^{\left(1-e^{\kappa t}\right) E_{\rm B}/T_{\rm in}}\,,
\end{align}
which allows to express
\begin{align}\label{EQ:naive_approx}
e^{\kappa t} &\approx \ln \left[e \left(\frac{{\mathfrak T}_{\rm in}}{\mathfrak T}\right)^{T_{\rm in}/E_{\rm B}}\right]\nn
{\mathfrak T}_{\rm in} &= -\frac{\Gamma T_{\rm in}}{\kappa E_{\rm B}} \exp\left\{-\frac{E_{\rm B}}{T_{\rm in}}\right\}
\,.
\end{align}
From that, we can conclude
\begin{align}
\tanh(2\beta({\mathfrak T})J) &= \tanh\left(\frac{2J}{T_{\rm in}} e^{\kappa t}\right)\nn
&\approx \tanh\left[\ln\left(e^{\frac{2J}{T_{\rm in}}} \left(\frac{{\mathfrak T}_{\rm in}}{\mathfrak T}\right)^{2J/E_{\rm B}}\right)\right]\nn
&= \frac{e^{4J/T_{\rm in}} \left(\frac{{\mathfrak T}_{\rm in}}{\mathfrak T}\right)^{4J/E_{\rm B}}-1}{e^{4J/T_{\rm in}} \left(\frac{{\mathfrak T}_{\rm in}}{\mathfrak T}\right)^{4J/E_{\rm B}}+1}\,.
\end{align}
For our parameters $4J \approx E_{\rm B}$ and $4J \gg T_{\rm in}$, we can thus simplify the source term
$\tanh(2\beta J) \to 1$ and the damping term as
$-2[1-\tanh(2\beta J)] \to (\frac{-4 {\mathfrak T}}{{\mathfrak T}_{\rm in}}) e^{-4 J/T_{\rm in}} $, 
which upon inserting ${\mathfrak T}_{\rm in} \approx - \frac{\Gamma T_{\rm in}}{\kappa E_{\rm B}} e^{-E_{\rm B}/T_{\rm in}}$ reproduces the simplified equation 
$\partial_{\mathfrak T} {\mathfrak C} = \lambda \mathfrak T \mathfrak C(\mathfrak T) + 1$
in the main text.
Its full analytic solution
\begin{align}
\mathfrak C (\mathfrak T) &= \mathfrak C_{\rm in} e^{\lambda(\mathfrak T^2 - \mathfrak T_{\rm in}^2)/2}\nn
&\qquad+ \sqrt{\frac{\pi}{2\lambda}} e^{\lambda \mathfrak T^2/2} \left[{\rm erf}\left(\frac{\sqrt{\lambda} \mathfrak T}{\sqrt{2}}\right)
- {\rm erf}\left(\frac{\sqrt{\lambda} \mathfrak T_{\rm in}}{\sqrt{2}}\right)\right]
\end{align}
then simplifies for $\mathfrak T\to0$ and $\sqrt{\lambda}\abs{\mathfrak T_{\rm in}} \gg 1$ to $\mathfrak C_{\rm freeze} \approx \sqrt{\pi/(2\lambda)}$.
Indeed, when $\mathfrak C_{\rm in}=0$ (which can be realized by preparing the spin system in a random state, i.e., without \new{prior equilibration}), one also observes 
agreement between analytic 1D curves and full 2D simulations for fast cooling sweeps, see Fig.~\ref{FIG:1DC0}.
\begin{figure}
\includegraphics[width=0.5\textwidth]{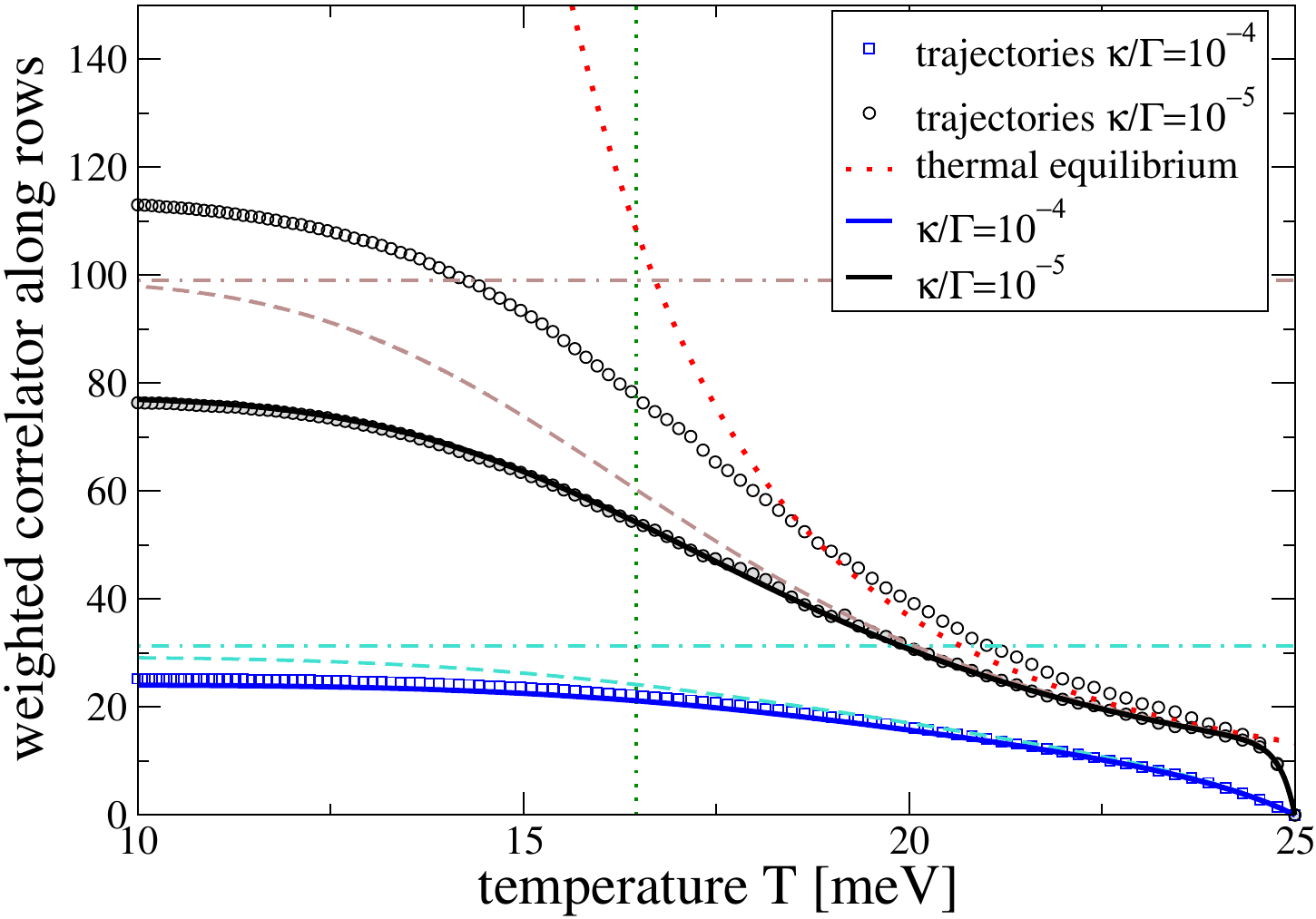}
\caption{
\new{Analogous to Fig.~1 
in color coding and other parameters, but without prior equilibration phase $\mathfrak C_{\rm in}=0$.
The same general features can be observed.
For slower cooling sweeps, full 2D trajectory simulations (hollow circle symbols) diverge from the numerical solution (solid black curve) of Eq.~(6).
When transverse couplings are turned off (filled circle symbols), the 1D Ising dynamics is faithfully reproduced.
}
}
\label{FIG:1DC0}
\end{figure}

\subsection{Inverse cooling sweep $T(t) = (\chi t)^{-1}$}

Even simpler considerations are possible for quenches where the inverse temperature is linear in time, i.e., $\beta(t) = \chi t$ or $T(t) \propto 1/T$, starting at infinite temperature
at $t=0$ and cooling down to zero temperature at $t\to\infty$.
Also in this case, the infinite laboratory time interval $t\in(0,\infty)$ is mapped to a finite interval of conformal time $\mathfrak T\in[\mathfrak T_{\rm in}, 0]$, but the relation between laboratory and conformal time
\begin{align}
\mathfrak T(t) = \mathfrak T_{\rm in} e^{-\chi E_{\rm B} t}\qquad:\qquad
\mathfrak T_{\rm in} = \frac{-\Gamma}{\chi E_{\rm B}}
\end{align}
\new{can be inverted exactly for $t$.}
For our parameters $E_{\rm B} \approx 4 J$, we can write
\begin{align}
\tanh(2 \beta J) &= \tanh\left[\ln\left(\frac{\mathfrak T_{\rm in}}{\mathfrak T}\right)^{2J/E_{\rm B}}\right]
=\frac{\left(\frac{\mathfrak T_{\rm in}}{\mathfrak T}\right)^{4J/E_{\rm B}}-1}{\left(\frac{\mathfrak T_{\rm in}}{\mathfrak T}\right)^{4J/E_{\rm B}}+1}\nn
&\to \frac{\mathfrak T_{\rm in}-\mathfrak T}{\mathfrak T_{\rm in}+\mathfrak T}\,,
\end{align}
which leads to the differential equation
\begin{align}
\partial_{\mathfrak T} \mathfrak C \approx -4 \frac{\mathfrak T}{\mathfrak T_{\rm in}+\mathfrak T} \mathfrak C + \frac{\mathfrak T_{\rm in}-\mathfrak T}{\mathfrak T_{\rm in}+\mathfrak T}\,.
\end{align}
Although this allows for an analytic solution, its long-term dynamics can be simplified further, as after a quick initial relaxation one has $\abs{\mathfrak T} \ll \abs{\mathfrak T_{\rm in}}$, after which the differential equation becomes
\begin{align}
\partial_{\mathfrak T} \mathfrak C \approx -4 \frac{\mathfrak T}{\mathfrak T_{\rm in}} \mathfrak C + 1\,,
\end{align}
which \new{-- analogous to exponential cooling -- also} admits for an analytic solution and equivalent scaling of the frozen correlation length
$\mathfrak C_{\rm freeze} \approx \sqrt{\pi \Gamma/(8 \chi E_{\rm B})}$ and furthermore agrees well with the corresponding numerical simulations, see Fig.~\ref{FIG:1Dold}.
}

\section{Appendix: 2D Ising model} 

For the anisotropic 2D Ising model -- compare Eq.~(1)
with $J_x\to J_\|$, $J_y\to J_\perp$, and $J_{\rm d}\to 0$ -- the energy difference upon flipping a 
single spin at site $(i,j)$ (i.e., $\sigma_{i,j}'=-\sigma_{i,j}$) is
\begin{align}
E_{\fk{\sigma}}-E_{\fk{\sigma}'} &= -2 J_\| \sigma_{i,j} (\sigma_{i-1,j}+\sigma_{i+1,j})\nn
&\quad - 2 J_\perp \sigma_{i,j} (\sigma_{i,j-1}+\sigma_{i,j+1}) \,,
\end{align}
such that the Glauber rates~(3)
lead to a quartic representation
\begin{align}
\frac{1}{e^{\beta(E_{\fk{\sigma}'}-E_{\fk{\sigma}})}+1} 
&= 
\frac{1}{2} - f_{\|} \sigma_{i,j} (\sigma_{i-1,j}+\sigma_{i+1,j})
\nn
&\quad-f_{\perp} \sigma_{i,j}(\sigma_{i,j-1}+\sigma_{i,j+1})\nn
&\quad+g_{\|} \sigma_{i,j}(\sigma_{i-1,j}+\sigma_{i+1,j})\sigma_{i,j-1}\sigma_{i,j+1}\nn
&\quad+g_{\perp} \sigma_{i,j}\sigma_{i-1,j}\sigma_{i+1,j}(\sigma_{i,j-1}+\sigma_{i,j+1})\,,
\end{align}
with the abbreviations
\begin{align}
f_{\|} &= \frac{[1+2 \cosh(4 \beta J_\|)+\cosh(4\beta J_\perp)]\tanh(2\beta J_\|)}{16 \cosh(2\beta J_\|+2\beta J_\perp)\cosh(2\beta J_\|-2\beta J_\perp)}\,,\nn
f_{\perp} &= \frac{[1+2 \cosh(4 \beta J_\perp)+\cosh(4\beta J_\|)]\tanh(2\beta J_\perp)}{16 \cosh(2\beta J_\| + 2\beta J_\perp)\cosh(2\beta J_\|-2\beta J_\perp)}\,,\nn
g_{\|} &= \frac{\sinh^2(2\beta J_\perp)\tanh(2\beta J_\|)}{4\cosh(4\beta J_\|) + 4\cosh(4\beta J_\perp)}\,,\nn
g_{\perp} &= \frac{\sinh^2(2\beta J_\|)\tanh(2\beta J_\perp)}{4\cosh(4\beta J_\|) + 4\cosh(4\beta J_\perp)}\,,
\end{align}
which shows that it will in general not be possible to obtain exact closed 
evolution equations, as e.g. first-order terms couple to three-point functions ($F_{ij}$ implements a flip of spin at $ij$)
\begin{align}
\partial_t \langle \sigma_{i,j}\rangle &= -2 \sum_{\fk{\sigma}} \sigma_{i,j} R_{\fk{\sigma} \to F_{ij} \fk{\sigma}} P_{\fk{\sigma}}\nn
&=\Gamma e^{-\beta E_{\rm B}} \Big[-\langle \sigma_{i,j}\rangle\nn
&\qquad+ 2 f_{\|} \langle \sigma_{i-1,j}+\sigma_{i+1,j}\rangle
+ 2 f_{\perp} \langle \sigma_{i,j-1}+\sigma_{i,j+1}\rangle\nn
&\qquad- 2 g_{\|} \langle (\sigma_{i-1,j}+\sigma_{i+1,j}) \sigma_{i,j-1} \sigma_{i,j+1}\rangle\nn
&\qquad- 2 g_{\perp} \langle \sigma_{i-1,j} \sigma_{i+1,j} (\sigma_{i,j-1}+\sigma_{i,j+1})\rangle\Big]\,,
\end{align}
and analogously two-point functions couple to four-point functions
\begin{align}
\partial_t \langle \sigma_{i,j} \sigma_{k,\ell} \rangle &= -2 \sum_{\fk{\sigma}} \sigma_{i,j} \sigma_{k,\ell} (R_{\fk{\sigma} \to F_{ij} \fk{\sigma}} + R_{\fk{\sigma} \to F_{k\ell} \fk{\sigma}}) P_{\fk{\sigma}}\nn
&= \Gamma e^{-\beta E_{\rm B}} \Big[-2\langle \sigma_{i,j} \sigma_{k,\ell}\rangle\nn
&\quad+2 f_{\|} \langle \sigma_{k,\ell}(\sigma_{i-1,j}+\sigma_{i+1,j})\rangle\nn
&\quad+2 f_{\|} \langle \sigma_{i,j} (\sigma_{k-1,\ell}+\sigma_{k+1,\ell})\rangle\nn
&\quad+2 f_{\perp} \langle \sigma_{k,\ell}(\sigma_{i,j-1}+\sigma_{i,j+1})\rangle\nn
&\quad+2 f_{\perp} \langle \sigma_{i,j} (\sigma_{k,\ell-1}+\sigma_{k,\ell+1})\rangle\nn
&\quad-2 g_{\|} \langle \sigma_{k,\ell}(\sigma_{i-1,j}+\sigma_{i+1,j})\sigma_{i,j-1} \sigma_{i,j+1}\rangle\nn
&\quad-2 g_{\|} \langle \sigma_{i,j}(\sigma_{k-1,\ell}+\sigma_{k+1,\ell})\sigma_{k,\ell-1} \sigma_{k,\ell+1}\rangle\nn
&\quad-2 g_{\perp} \langle \sigma_{k,\ell}\sigma_{i-1,j}\sigma_{i+1,j}(\sigma_{i,j-1}+\sigma_{i,j+1})\rangle\nn
&\quad-2 g_{\perp} \langle \sigma_{i,j}\sigma_{k-1,\ell}\sigma_{k+1,\ell}(\sigma_{k,\ell-1}+\sigma_{k,\ell+1})\rangle\Big]\,,
\end{align}
and so on.
The above hierarchy can be closed at some order by 
mean-field-type approximations, leading in general to nonlinear equations, which however is expected to become inaccurate near the critical point.

If we would keep only two-point correlators and neglect all three-point 
and higher correlations, the above equation would 
again become a diffusion-dissipation equation (in the continuum limit), 
but with an anisotropic diffusion kernel.
The damping term $\gamma$ would then vanish at a mean-field approximation to the critical temperature 
$T_{\rm crit}$ of the 2D Ising model~(1).
However, while the restriction to two-point correlations is exact for 
the 1D Ising model, it is only an approximation for the 2D case. 
Moreover, this approximation is expected to become quite inaccurate 
near the critical point. 

In the strongly anisotropic scenario relevant for our considerations, noting that $g_{\perp}=\ord(J_\perp)$ and considering the strong correlations along rows
\new{and sweeps for which the initial correlations across rows are negligible}, we can approximate terms like
$\langle \sigma_{k,\ell} \sigma_{i-1,j}\sigma_{i+1,j} \sigma_{i,j\pm 1}\rangle \approx \langle\sigma_{k,\ell} \sigma_{i,j\pm 1}\rangle \langle \sigma_{i-1,j}\sigma_{i+1,j}\rangle
\approx \langle \sigma_{k,\ell} \sigma_{i,j\pm 1}\rangle \new{c_2(t)}$,
\new{where $c_2$ is the time-dependent solution of Eq.~(5).}
Taking $k=i+a$ and $\ell=j+b$, and exploiting translational invariance, we then obtain for the correlator (further using that $g_{\|} = \ord(J_\perp^2)$),
\begin{align}\label{EQ:next}
\partial_t c_{a,b} &= \Gamma e^{-\beta E_{\rm B}}\Big[-2 c_{a,b} + 4 f_{\|} (c_{a-1,b}+c_{a+1,b})\nn
&\qquad + 4 [f_\perp-g_\perp \new{c_2})] (c_{a,b-1}+c_{a,b+1})\Big]\nn
&\qquad + \ord(J_\perp^2)\,.
\end{align}
Furthermore inserting $4f_{\|} = \tanh(2\beta J_\|) + \ord(J_\perp^2)$ and $4 [f_\perp-g_\perp \tanh^2(\beta J_\|)]=2\beta J_\perp/\cosh(2\beta J_{\|}) + \ord(J_\perp^2)$, \new{$c_2 \to \tanh^2(\beta J_\|)$}, and employing the mapping to the conformal time as before, we recover Eq.~(8)
in the main text.
An approximation to the critical temperature is then obtained from solving $1=4 f_{\|} + 4 [f_\perp-g_\perp \tanh^2(\beta_{\rm crit} J_\|)]$.

If instead we had neglected all four-point correlators completely, we would obtain the above equation with $g_\perp \to 0$, which with $4 f_\perp = \beta J_\perp [1+\cosh^{-2}(2\beta J_\|)] + \ord(J_\perp^2)$ would lead to a larger diffusion constant along the weakly-coupled direction.
Also the equation for the approximate critical point $1=4 f_{\|} + 4 f_\perp$ would yield a less accurate result in this case.

\new{For our highly anisotropic couplings, we may approximate the initial condition by the 1D Ising equilibrium solution
\begin{align}
c_{a,b}(\tau_0) = \tanh^a(\beta_0 J_{\|}) \delta_{b,0} + \ord(J_\perp)\,,
\end{align}
from which we may also conclude that $c_{a,1}(\tau) = \ord(J_\perp)$ and $c_{a,2}(\tau)=\ord(J_\perp^2)$, such that we may further approximate~\eqref{EQ:next} in conformal time as
\begin{align}
\partial_{\mathfrak T} c_{a\neq 0,0} &= -2 c_{a,0} + 4 f_{\|}(\mathfrak T) (c_{a-1,0}+c_{a+1,0})+ \ord(J_\perp^2)\,,\nn
\partial_{\mathfrak T} c_{0,0} &= 0\,,\nn
\partial_{\mathfrak T} c_{a,1} &= -2 c_{a,1} + 4 f_{\|}(\mathfrak T) (c_{a-1,1}+c_{a+1,1})\nn
&\qquad+ 4 [f_\perp(\mathfrak T)-g_\perp(\mathfrak T) c_2(\mathfrak T)] c_{a,0} + \ord(J_\perp^2)\,.
\end{align}
Inserting the asymptotics of the $f_\nu(\tau)$ and $g_\nu(\tau)$ for small $J_\perp$, this translates into
\begin{align}\label{EQ:solved_system}
\partial_{\mathfrak T}  c_{a\neq 0,0} &= -2 c_{a,0} + \tanh(2\beta(\mathfrak T)J_{\|}) (c_{a-1,0}+c_{a+1,0})\nn
&\qquad+ \ord(J_\perp^2)\,,\nn
\partial_{\mathfrak T}  c_{0,0} &= 0\,,\nn
\partial_{\mathfrak T}  c_{a,1} &= -2 c_{a,1} + \tanh(2\beta(\mathfrak T)J_{\|}) (c_{a-1,1}+c_{a+1,1})\nn
&\qquad+ \beta(\mathfrak T) J_\perp [2-\tanh^2(2\beta(\mathfrak T)J_{\|})(1+c_2(\mathfrak T))]c_{a,0}\nn
&\qquad + \ord(J_\perp^2)\,.
\end{align}
The first of these equations has exactly the same structure and initial condition as in the 1D case~(5)
and thus has the same solution $c_{a,0}(\mathfrak T)\approx c_a(\mathfrak T)$ for $a\in\{1,2,\ldots,N-1\}$.
The last equation has also the same structure but a different inhomogoneity, which also allows for a rather straighforward solution.

From 
the 2D simulations we can easily extract the individual correlators.
However, as these are subject to finite-size fluctuations, we found it simpler to construct observables from the sum of squared correlators along and across the rows, 
which can be extracted from the Fourier-transformed lattice.
One should keep in mind that the latter includes the on-site correlator $C_{0,0}$ and due to the periodic boundary condition also involves a double-counting of all other correlators.
Therefore, we compare 
$\sum_a C_{a,0}^2$ from the D simulation with $1 + 2 \sum_a c_{a,0}^2(\mathfrak T)\approx 1+2 \sum_a c_a^2(\mathfrak T)$ from the correlator equations
and accordingly 
$\sum_b C_{0,b}^2$ with $1+2 \sum_b c_{0,b}^2(\mathfrak T) \approx 1+2 c_{0,1}^2(\mathfrak T)$.

Furthermore, to perform quantitative comparisons, it is beneficial to improve on the inversion of the conformal time in Eq.~\eqref{EQ:naive_approx}.

The first observation is that the initial condition for the approximate conformal time~\eqref{EQ:naive_approx} does not match the exact initial condition of~\eqref{EQ:conf_time}.
One can easily correct for this by simply inserting the proper initial condition, which upgrades~\eqref{EQ:naive_approx} to
\begin{align}\label{EQ:exp_approx}
e^{\kappa t} &\approx \ln \left[e \left(\frac{{\mathfrak T}_{\rm in}^{\rm ex}}{\mathfrak T}\right)^{T_{\rm in}/E_{\rm B}}\right]\nn
{\mathfrak T}_{\rm in}^{\rm ex} &= {\rm Ei}\left(-e^{\kappa t} \frac{E_{\rm B}}{T_{\rm in}} \right) \frac{\Gamma}{\kappa}\,, 
\end{align}
and to which we will refer as {\em exponential approximation}.

Second, we see that in the asymptotic expansion of the exponential integral, it is well possible to include one more term
$\mathfrak T(t) \approx \frac{\Gamma}{\kappa} {\rm Ei}\left(-\frac{E_{\rm B}}{T_{\rm in}}\right)\exp\left\{\frac{E_{\rm B}}{T_{\rm in}}\left(1-e^{\kappa t}\right)\right\} e^{-\kappa t} \equiv \mathfrak T_{\rm in}^{\rm ex} \exp\left\{\frac{E_{\rm B}}{T_{\rm in}}\left(1-e^{\kappa t}\right)\right\} e^{-\kappa t}$, which eventually yields
\begin{align}\label{EQ:plog_approx}
e^{\kappa t} &\approx \frac{T_{\rm in}}{E_{\rm B}} {\rm W}\left(\frac{\mathfrak T_{\rm in}^{\rm ex}}{\mathfrak T}\frac{E_{\rm B}}{T_{\rm in}} e^{E_{\rm B}/T_{\rm in}}\right)\,,
\end{align}
which also uses the exact initial condition and 
where $W(x)$ is the productlog function, defined as the solution of  $x=w e^w$ for $w$.
We will refer to this as the {\em productlog approximation}.

Third, we may also solve Eq.~\eqref{EQ:conf_time} numerically for the laboratory time, which we denote as {\em numerical inversion}.

The results are shown in Fig.~\ref{FIG:correlators}, where one can observe a convincing agreement between 2D simulations (symbols) and the results from the correlator equations (curves).
\begin{figure}
\includegraphics[width=0.5\textwidth]{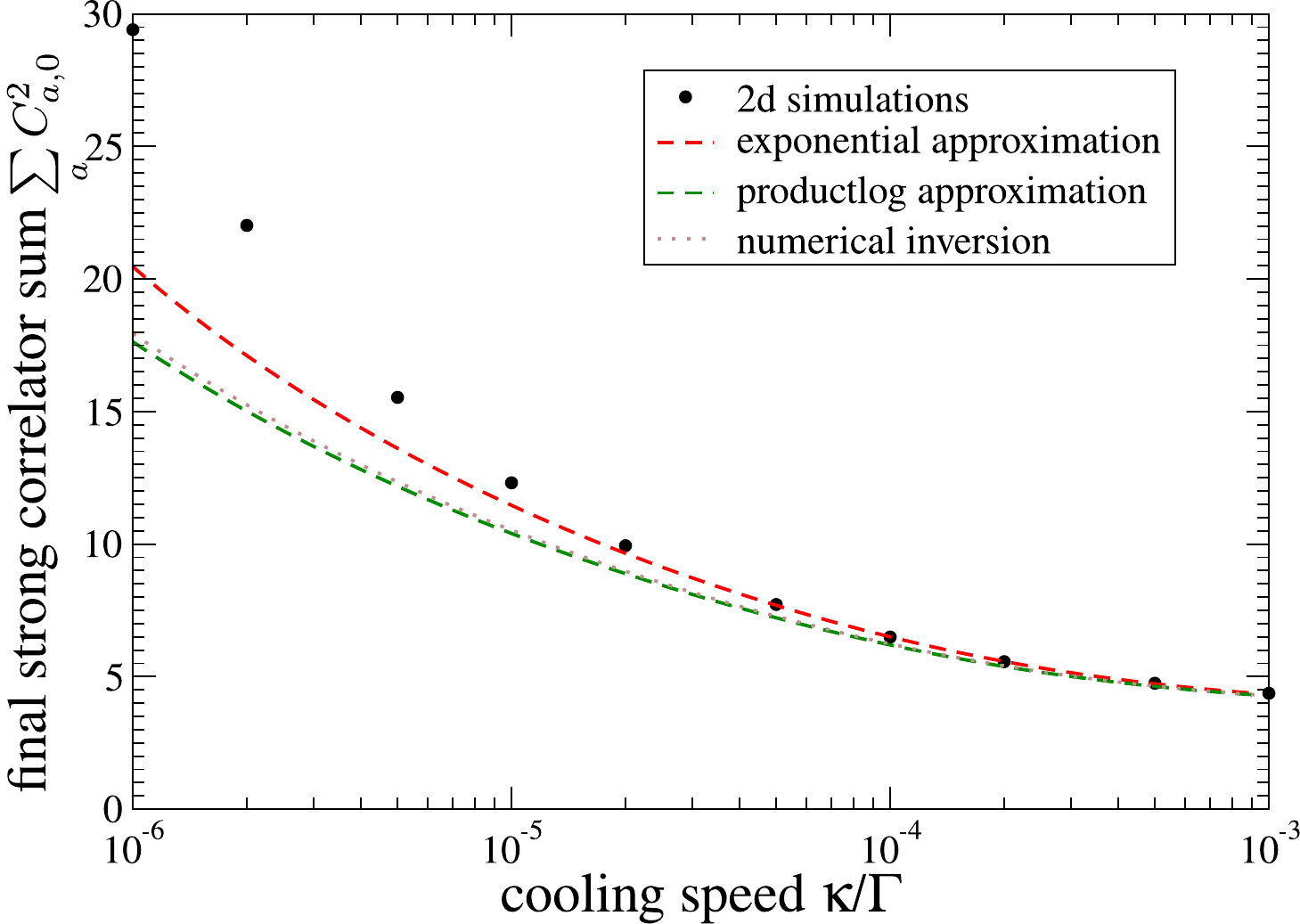}\\
\includegraphics[width=0.5\textwidth]{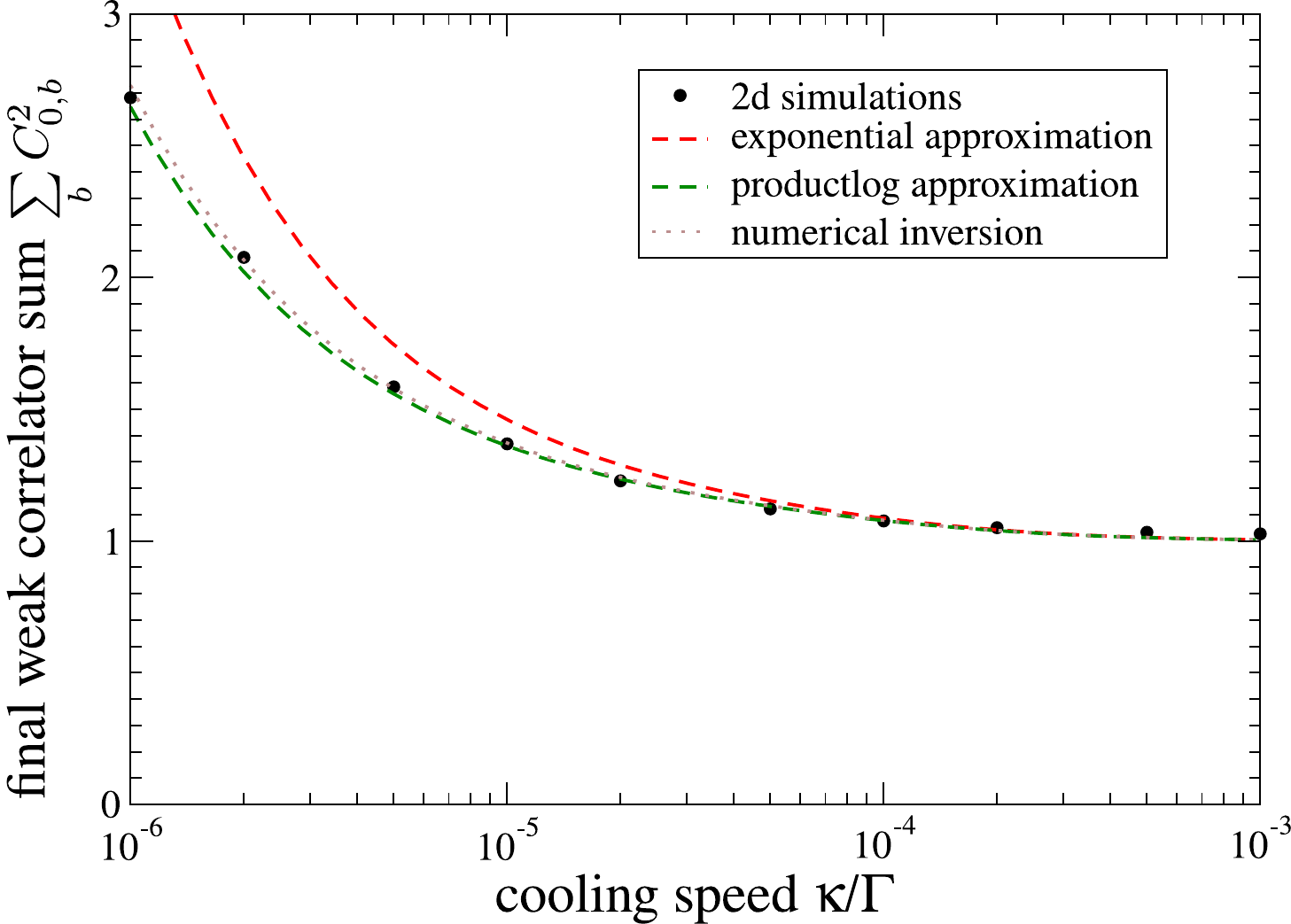}
\caption{\label{FIG:correlators}
\new{Plot of the sum of squared correlators in strongly coupled (along rows, top) and weakly coupled (across rows, bottom) directions. 
The curves are the equivalent quantities extracted from the $2D$ correlator equations~\eqref{EQ:solved_system} for approximations~\eqref{EQ:exp_approx} and~\eqref{EQ:plog_approx} on the conformal time mapping or for the exact numerical inversion of~\eqref{EQ:conf_time}, whereas symbols are extracted from the numerical simulations. For fast quenches (large $\kappa$), convincing agreement is found.}
}
\end{figure}
Furthermore, one can see that for our parameters, the productlog approximation is already matching the numerical inversion quite well, and the differences between these curves and the 2D numerical simulations are due to the approximations performed in the correlator equations.
Nevertheless, already 
the exponential approximation (dashed red) captures the scaling of the correlators.
}
\section{Appendix: Numerical Methods} 

\paragraph{Stochastic propagation}

To solve a rate equation of the form
\begin{align}
\dot{\f{P}} = {\cal L}(t) \f{P}\,,
\end{align}
with probability vector $\f{P}$ and time-dependent rate matrix
${\cal L}(t) = {\cal L}_0(t) + {\cal L}_1(t)$ (with diagonal part ${\cal L}_0(t)$ and an off-diagonal part ${\cal L}_1(t)$) via trajectories, 
we employ a Dyson-type series for the full propagator
\begin{align}
{\cal P}(t,0) &= {\cal P}_0(t,0) + \int_0^t dt_1 {\cal P}_0(t,t_1) {\cal L}_1(t_1) {\cal P}_0(t_1,0)\nn
&\qquad+ \int_0^t dt_2 \int_0^{t_2} dt_1 {\cal P}_0(t,t_2) {\cal L}_1(t_2) {\cal P}_0(t_2,t_1)\times\nn
&\qquad\qquad\times {\cal L}_1(t_1) {\cal P}_0(t_1,0) + \ldots
\end{align}
As ${\cal L}_0(t)$ is a diagonal matrix, it commutes with itself at different times, and  we can compute the associated propagator simply via
\begin{align}
{\cal P}_0(t_2,t_1) = \exp\left\{\int_{t_1}^{t_2} {\cal L}_0(t') dt'\right\}\,.
\end{align}
Therefore, assuming that at time $t$ we are in state $\f{\sigma}$, the associated probability to remain in that state up to time $t+\tau$ is
\begin{align}
P_{\rm stay}^{\fk{\sigma}}(t,\tau) = \exp\left\{-\sum_{\fk{\sigma'} \neq \fk{\sigma}} \int_t^{t+\tau} R_{\fk{\sigma}\to\fk{\sigma'}}(t') dt'\right\}\,.
\end{align}
The waiting time distribution for the first jump to a different state is then
\begin{align}
w^{\fk{\sigma}}(t,\tau) = \frac{d}{d\tau} \left[1-P_{\rm stay}^{\fk{\sigma}}(t,\tau)\right]\,.
\end{align}
To obtain accordingly distributed waiting times from uniformly distributed random numbers $r\in[0,1]$ we thus equate those with the cumulative probability of jumping
$r = 1 - \exp\left\{-\sum_{\fk{\sigma'} \neq \fk{\sigma}} \int_t^{t+\tau} R_{\fk{\sigma}\to\fk{\sigma'}}(t') dt'\right\}$, such that we solve the equation
\begin{align}
\ln (1-r) = - \sum_{\fk{\sigma'} \neq \fk{\sigma}} \int_t^{t+\tau} R_{\fk{\sigma}\to\fk{\sigma'}}(t') dt'
\end{align}
for the waiting time $\tau$.
This can be achieved by Newton-Raphson iteration, where a suitable initial guess can be obtained by assuming that rates are roughly constant over the waiting time
\begin{align}
\tau_0 = -\frac{\ln(1-r)}{\sum\limits_{\fk{\sigma'} \neq \fk{\sigma}} R_{\fk{\sigma}\to\fk{\sigma'}}(t)}\,.
\end{align}

\paragraph{Extraction of correlation lengths}

Denoting the (fast) Fourier-transformed grid for an $N_x \times N_y$ lattice as
\begin{align}
\tilde \sigma_{k\ell} &= \sum_{ij} \sigma_{ij} e^{-2\pi\ii (i-1)(k-1)/N_x} e^{-2\pi\ii (j-1)(\ell-1)/N_y}\,,
\end{align}
it follows that the quantities
\begin{align}
\sum_k \abs{\tilde\sigma_{k\ell}}^2 &= N_x \sum_q  \left[\sum_{ij} \sigma_{ij} \sigma_{i,j+q}\right] e^{+2\pi\ii q(\ell-1)/N_y}\,,\nn
\sum_\ell \abs{\tilde\sigma_{k\ell}}^2 &= N_y \sum_q \left[\sum_{ij} \sigma_{ij} \sigma_{i+q,j}\right] e^{+2\pi\ii q(k-1)/N_x}\,,
\end{align}
are Fourier transforms of the convolutions $\sum_{ij} \sigma_{ij} \sigma_{i,j+q}$ and $\sum_{ij} \sigma_{ij} \sigma_{i+q,j}$.
By ensemble-averaging (runs can and have been performed in parallel), we thus have
\begin{align}
\sum_k \expval{\abs{\tilde\sigma_{k\ell}}^2} &= N_x \sum_q \left[\sum_{ij} \expval{\sigma_{ij} \sigma_{i,j+q}}\right] e^{+2\pi\ii q(\ell-1)/N_y}\,,\nn
\sum_\ell \expval{\abs{\tilde\sigma_{k\ell}}^2} &= N_y \sum_q \left[\sum_{ij} \expval{\sigma_{ij} \sigma_{i+q,j}}\right] e^{+2\pi\ii q(k-1)/N_x}\,.
\end{align}
Accordingly, when the expectation values decay like $\expval{\sigma_{ij} \sigma_{i+q,j}} \propto e^{-\abs{q}/\xi_\|}$ and 
$\expval{\sigma_{ij} \sigma_{i,j+q}} \propto e^{-\abs{q}/\xi_\perp}$ with correlation lengths $\xi_\|$ and $\xi_\perp$, it follows that we can obtain the latter from the inverse widths
of $\sum_\ell \abs{\tilde\sigma_{k\ell}}^2$ and $\sum_k \abs{\tilde\sigma_{k\ell}}^2$, respectively.
The subtraction of an average magnetization has no impact on our simulations.

\new{
The above relations also show that the correlators (and also the sum of their squares) can be extracted from the Fourier transformed lattices, e.g.
\begin{align}
C_{a,0} &= \frac{1}{N_x N_y} \sum_{ij}\expval{\sigma_{i,j}\sigma_{i+a,j}}\nn
&= \frac{1}{N_x^2 N_y^2} \sum_k \expval{\sum_\ell \abs{\tilde\sigma_{k,\ell}}^2} e^{-2\pi\ii(k-1)/N_x}\,,
\end{align}
and analogously for the weakly coupled direction.

For the extraction of the weighted sum of correlations $\mathfrak C(\mathfrak T)$, it should be noted that Fourier transforms for finite-size simulations
have considerable tails, such that for the numerically stable computation of the weighted correlators one has to use a cutoff length $N_{\rm cut}$
\begin{align}
\mathfrak{C} &= \sum_{a=1}^{N_{\rm cut}} \frac{a}{N_x N_y} \sum_{ij} \frac{1}{2}\left[\expval{\sigma_{i,j} \sigma_{i+a,j}}+\expval{\sigma_{i,j}\sigma_{i+N_x-a,j}}\right]\nn
&= \sum_{a=1}^{N_{\rm cut}} \frac{a}{N_x^2 N_y^2} \sum_{k\ell} \expval{\abs{\tilde \sigma_{k\ell}}^2} \cos\left(2\pi\frac{a(k-1)}{N_x}\right)\,,
\end{align}
and analogously for the perpendicular direction.
This cutoff length marks the distance above which two sites are essentially uncorrelated.
As the correlators for uncorrelated sites may still exhibit fluctuations $\Delta C_{a,b}^{\rm uncorr} \approx \frac{1}{\sqrt{N_x N_y N_{\rm trj}}}$, we determined the cutoff length by
the condition
$C_{N_{\rm cut},0} \approx \frac{5}{\sqrt{N_x N_y N_{\rm trj}}}$ and analogously for the perpendicular direction.
}

\end{document}